\documentclass[12pt]{article}
\usepackage{amsmath,amssymb,epsfig}

\usepackage{color}
\input{colordvi.tex}
\def\unit{{\relax{\rm 1\kern-.26em I}}}

\newcommand{\tr}{{\rm Tr}}

\makeatletter

\renewcommand\section{\@startsection {section}{1}{\z@}%
                                   {-3.5ex \@plus -1ex \@minus -.2ex}%
                                   {2.3ex \@plus.2ex}%
                                   {\normalfont\large\bfseries}}

\renewcommand\subsection{\@startsection{subsection}{2}{\z@}%
                                     {-3.25ex\@plus -1ex \@minus -.2ex}%
                                     {1.5ex \@plus .2ex}%
                                     {\normalfont\normalsize\bfseries}}

\newcount\hour \newcount\minute
\hour=\time \divide \hour by 60
\minute=\time
\def\now{%
\ifnum \hour<13
  \ifnum \hour=0 \advance \hour by 12 \number\hour:\else \number\hour:\fi%
     \ifnum \minute<10 0\fi%
     \number\minute%
\ A.M.%
\else \advance \hour by -12 \number\hour:%
  \ifnum \minute<10 0\fi%
  \number\minute%
  \ P.M.%
\fi%
}

\makeatother

\begin{document}

\baselineskip=18pt  
\numberwithin{equation}{section}  
\allowdisplaybreaks  



%
%


\thispagestyle{empty}

\vspace*{-2cm}
\begin{flushright}
{\tt YGHP-12-51}, {\tt IFUP-TH/2012-22}
\end{flushright}

\begin{flushright}
\end{flushright}

\begin{center}

\vspace{-.5cm}

\vspace{0.5cm}
{\bf\Large Brane Realization of Nambu Monopoles }
\vspace*{0.5cm}

{\bf\Large  and Electroweak Strings}
\vspace*{1.5cm}

{\bf
Minoru Eto$^{1}$, Kenichi Konishi$^{2,3}$, Muneto Nitta$^{4}$ and Yutaka Ookouchi$^{5}$}
\vspace*{0.5cm}
 
\vspace*{0.5cm}

$^{1}${\it {Department of Physics, Yamagata University, Yamagata 990-8560, Japan}}
\vspace{0.1cm}

$^2$ {\it Department of Physics, ``E. Fermi'',  University of Pisa,   \\  Largo Pontecorvo, 3,  Pisa  56127, Italy }\\
\vspace{0.1cm}

$^{3}${\it  INFN, Sezione di Pisa,        Largo Pontecorvo, 3,  Pisa  56127, Italy }\\
\vspace{0.1cm}

\vspace{0.1cm}

$^{4}${\it Department of Physics, and Research and Education Center for Natural 
Sciences, Keio University, Hiyoshi 4-1-1, Yokohama, Kanagawa 223-8521, Japan}\\
\vspace{0.1cm}

$^{5}${\it The Hakubi Center for Advanced Research \& Department of Physics, \\ Kyoto University, Kyoto 606-8502, Japan }\\

\end{center}

\vspace{1cm} \centerline{\bf Abstract} \vspace*{0.5cm}

In the standard model, the electroweak $Z$-string can end on a Nambu monopole, 
whose mass is calculated to be 689 GeV from the current precise experimental data 
 assuming the new particle with mass 125 GeV to be the Higgs boson.
We study an extension of the standard model with
additional singlet and triplet Higgs fields
in the framework of ${\cal N}=1$ supersymmetric field theory 
by using a D-brane configuration in type IIA string theory. 
We construct a D-brane configuration describing the electroweak symmetry breaking, 
and find a single D2-brane configuration describing a $Z$-string and 
a Nambu monopole attached by a $Z$-string
in the standard model without an adjoint Higgs field.
We further find a single D2-brane configuration describing  
a composite of a 't Hooft-Polyakov monopole and a Nambu monopole attached by a $Z$-string 
in an extended standard model with an adjoint Higgs field.  
We compute the binding energy of the 't Hooft-Polyakov and Nambu monopoles
by solving a minimal surface area of a D2-brane.

\newpage
\setcounter{page}{1} 




\section{Introduction}

After Nambu imported the concept of spontaneous symmetry breaking (SSB) 
from the theory of superconductivity to high energy physics, 
it is now one of the most important concepts 
in modern physics from condensed matter physics to high energy physics.
In fact, SSB is a key ingredient of the unification of fundamental forces; 
one of the greatest achievements would be the electro-weak unification, 
in which the electromagnetism and the weak interaction is unified. 
Recently a new particle with the mass 125 GeV, 
which is most likely to be  identified with the Higgs boson, 
has been found \cite{Higgs-mass}, which makes the standard model being completed.

On the other hand, 
one of the inevitable consequences of SSB is the Kibble-Zurek mechanism 
\cite{Kibble:1976sj, Zurek:1985}
that creates topological defects at the phase transitions. 
While that mechanism might produce various kind of (non-)topological defects 
in the early Universe where the electroweak symmetry breaking occurs, 
it has been experimentally  verified in 
symmetry breakings in various condensed matter systems \cite{KZ-cond}.  
Topological and non-topological defects in the electroweak theory 
have been studied extensively for decades, 
such as sphalerons \cite{Manton:1983nd}, 
monopoles \cite{Nambu,Vachaspati:1994xe,Cho:1996qd}, and 
electroweak strings \cite{Nambu,Vachaspati:1992fi,JamesII,JamesI,Goodband:1995he,AVreview}, 
see \cite{AVreview,Vachaspati:1994xc,Yang:1997qz} as a review.
Among those, the electroweak strings were studied extensively, 
which are 
also called $Z$-strings because the most stable string carries the magnetic fluxes of 
the $Z$-boson. 
In particular, the (in)stability of the electroweak strings was studied in detail;  
although they are stable in a tiny region of the parameter space, 
they are unstable in the rest of wide range of parameter space 
\cite{JamesII,JamesI, Goodband:1995he,AVreview}. 
With the current experimental data, there are no stable electroweak strings within 
the minimum setup of the standard model. 
The electroweak strings become semi-local strings \cite{VA, Gibbons, Hind, AVreview} 
for vanishing weak gauge coupling of 
$SU(2)_{\rm W}$, in which the $SU(2)$ symmetry becomes global.
The semi-local strings are marginally stable which lie in the stable region of the parameter space 
of the electroweak strings. 
It was also discussed that the existence of fermion (quark or lepton) zero modes trapped 
inside the vortex core changes the (in)stability of the 
electroweak strings \cite{Vachaspati:1992mk}, but it does not seem to be conclusive.
However, even if the electroweak strings are unstable,  
they might be created at the electroweak phase transition 
and would play various roles in the early Universe, 
such as baryogenesis \cite{Brandenberger:1994bx}
and generation of primordial magnetic fields \cite{Vachaspati:1994xc,Poltis:2010yu}.

As found by Nambu \cite{Nambu} in his seminal paper, 
the electroweak string can terminate on a magnetic monopole, 
which is called the Nambu monopole. 
The mass of the Nambu monopole can be calculated to be 689 GeV from 
the current precise experimental data 
 assuming the new particle with mass 125 GeV to be the Higgs boson \cite{Higgs-mass}.

Therefore, the electroweak strings can decay even in the stable parameter region,
because a pair of a monopole and an anti-monopole can be created on the string 
by quantum mechanical tunneling or thermal fluctuations 
even in the presence of the potential barrier. 
The electroweak strings are at most metastable in this sense.
The Nambu monopole does not exist alone since the monopole carries 
the $Z$-flux which is confined into the flux tube in the spontaneously broken $U(1)_Z$ group. 
This gives a prototype of monopole-string composites, 
which describe, in the context of QCD, a dual superconductor picture 
of the quark confinement \cite{Nambu:1974zg}. 
Non-Abelian extension of such vortex-monopole composites are now studied extensively 
in supersymmetric QCD \cite{Auzzi:2003fs,Auzzi:2003em}. 
In the early Universe, there might exist an era that 
the electroweak strings and the Nambu monopoles are (meta)stable.
One possible way to make them (meta)stable is to add singlet and  adjoint Higgs fields
to the electroweak sector \cite{KeVa}.\footnote{
Many extensions of the minimal Higgs sector in the Standard Model were proposed in
order to explain several phenomena beyond the Standard Model such as tiny neutrino mass,
dark matter and baryon asymmetry of the Universe. For example, it is known that a triplet Higgs field
can generate neutrino masses at tree level, which is the so-called type-II seesaw model.
See, for example, \cite{Aoki:2012yt} and references therein.
} 
In that case, decay rate of the quantum mechanical decay by the monopole creations 
is very important. 

In this paper, we study dynamics of the Nambu monopoles and the electroweak strings
by using D-branes in superstring theory. 
To this end, we propose an 
${\cal N}=1$ supersymmetric extension of the Higgs sector in the Standard Model, 
which can be obtained from ${\cal N}=2$ supersymmetric theory by a deformation. 
Such a model can be realized as a low-energy effective action on
D-branes in superstring theory.
Our model contains the adjoint Higgs field as in \cite{KeVa}.  
We show in the context of type IIA string that the electroweak string ending on the Nambu monopole 
can be realized by a single D2-brane in the 4 dimensional Minkowski spacetime which bends into the internal space. Geometric understanding gives us a uniform manner to realize  various composites of the monopole and the string. 
A D-brane description of a Bogomol'nyi-Prasad-Sommerfield (BPS) monopole-vortex composite 
by a single D2 brane was studied before 
in ${\cal N}=2$ supersymmetric gauge theories 
\cite{Hanany:2004ea,Auzzi:2004yg}.
In particular, a binding energy between a monopole and a vortex 
was studied in Ref.~\cite{Auzzi:2004yg}.
One of the interesting consequences of our model is the existence of a composite of the `t Hooft-Polyakov and Nambu monopoles, which can be also understood by a single D2-brane. 
The calculation of binding energy of the composite monopole from the field theory side is a slightly involved task because it is a non-BPS object. On the other hand, from the geometric perspective, 
we calculate the binding energy of the 't Hooft-Polyakov-Nambu monopole by solving a Dirichlet problem of finding the two-dimensional minimal surface, which is one of advantages of geometric picture by D-branes. We will show some  examples of binding energy of the two monopoles. 
While the binding energy between a monopole and a vortex 
was studied in Ref.~\cite{Auzzi:2004yg}, we concentrate on the binding energy between the `t Hooft-Polyakov and the Nambu monopoles in this paper.

The structure of this paper is the following.
In section \ref{sec:field-theory}, we provide an ${\cal N}=1$ supersymmetric extension of the Higgs sector of the Standard Model. We study electroweak string and monopoles. Section \ref{sec:geometric} is devoted for studying these field theoretical solitons in the type IIA superstring theory. We will find that a single D2 brane plays an important role for understanding both the electroweak string
and the monopoles. In section \ref{sec:summary}, we will give concluding remarks and possible future directions.

\section{Field theory analysis}\label{sec:field-theory}

\subsection{Supersymmetric Higgs sector}

The ${\cal N}=1$ supersymmetric model we will focus on in this paper is as follows: Gauge symmetry is the electroweak symmetry\footnote{One may wonder that gauge group is $U(2)$ rather than $SU(2)$. However, in this case, by choosing the linear combination of two $U(1)$'s  appropriately, one can make that all the field in the theory carry only one $U(1)$ charge. Thus, since the other $U(1)$ symmetry is decoupled, we can omit the symmetry.} $U(1)_{\rm Y}\times SU(2)_{\rm W}$. To make the Higgs sector supersymmetric we introduce two Higgs doublets, $H, \tilde{H}$. In addition, we include a 2 by 2 hermitian matrix field $\Phi$ which is an adjoint (traceless part) plus singlet (trace part) representation of $SU(2)_{\rm W}$. All matter contents and charge assignments for local and global symmetries are summarized in Table \ref{matter}.
\begin{table}[htbp]
\begin{center}
\begin{tabular}{c|c|c|c}
\hline\hline
 &  $SU(2)_{\rm W}$ & $U(1)_{\rm Y}$ & $U(1)_R$ \\
\hline
$\Phi $& ${\bf adj+1}$ & $0$ & $2$ \\
$H$ & ${\bf 2}$ & $1$ & $0$ \\
$\tilde{H}$ & ${\bf 2}$ & $-1$ & $0$ \\
\hline
\end{tabular}
\end{center}
\caption{Matter contents and symmetries}
\label{matter}
\end{table}

We consider the supersymmetric model because supersymmetry provides us better controllability. Furthermore, it gives us a powerful geometric realization such as intersecting D-branes. As has been demonstrated in various context \cite{GiveonKutasov}, D-brane realizations give us simple insights of dynamics of gauge theory. Various information can be seen as geometric quantities in the internal space of string theory.

The key consequences of including the extra-field\footnote{See \cite{KeVa} for an early attempt to include an $SU(2)$ adjoint field. A crucial difference from our model is the existence of a singlet. Because of the singlet, we are allowed to take the vacuum expectation value (VEV) of $\langle \Phi\rangle ={\rm diag} (0,x)$. Thus, by changing $x$, as we will see in the next section, one can see the interpolation between an Abrikosov-Neilsen-Olesen(ANO) string and a semi-local string. }  $\Phi$ are twofold: First, the vacuum expectation value of the adjoint field yields symmetry breaking of $SU(2)_{\rm W} \to U(1)_{\rm W}$ and creates the `t Hooft-Polyakov monopole. As we will see below, this `t Hooft-Polyakov monopole constructs a bound state with the Nambu monopole, which is a fascinating aspect of our extended model. 
Second, in our non-minimal model, we can make the electroweak string stable, depending on the vacuum expectation value of the field. As mentioned in the Introduction, within the Standard Model, there is no stable electroweak string. Possibility of a stable electroweak string may have strong impact in the early age of the Universe. An interesting fact is that this stability is closely related to interpolation between an Abrikosov-Neilsen-Olesen (ANO) string \cite{Abrikosov:1956sx} and a semi-local string.

To realize electroweak symmetry breaking in our model, we introduce the following simple superpotential terms:
\begin{equation}
W= \sqrt{\beta} g \left(H \Phi  \tilde{H} - \mu^2  \tr [\Phi ] \right),
\end{equation}
where $g$ is the $SU(2)_{\rm W}$ gauge coupling as if there exists ${\cal N}=2$ supersymmetry instead of ${\cal N}=1$ supersymmetry \cite{Kutasov1}. We assume that the  K\"ahler potential for all fields are canonical. 
The kinetic terms are given by
\begin{eqnarray}
{\cal L}_Y = - \frac{1}{4}Y_{\mu\nu} Y^{\mu\nu},\quad
{\cal L}_W = - \frac{1}{4}W^a_{\mu\nu} W^{a\mu\nu},\qquad\qquad\\
{\cal L}_H = {\cal D}_\mu H {\cal D}^\mu H^\dagger,\quad
{\cal L}_{\tilde H} = {\cal D}_\mu \tilde H^\dagger {\cal D}^\mu \tilde H,\quad
{\cal L}_{\Phi} = \frac{1}{2}{\cal D}_\mu \Phi^\alpha {\cal D}^\mu \Phi^\alpha,
\end{eqnarray}
where $a=1,2,3$ and $\alpha = 0,1,2,3$.
The field strengths and the covariant derivatives are defined by
\begin{eqnarray}
Y_{\mu\nu} = \partial_\mu Y_\nu - \partial_\nu Y_\mu,\quad
W_{\mu\nu} = \partial_\mu W_\nu - \partial_\nu W_\mu + i g \left[W_\mu, W_\nu\right],\quad
{\cal D}_\mu \Phi = \partial_\mu \Phi + i g \left[W_\mu, \Phi\right],\\
{\cal D}_\mu H = \partial_\mu H - i gHW_\mu - i\frac{g'}{2}Y_\mu H,\quad
{\cal D}_\mu \tilde H^\dagger = \partial_\mu \tilde H^\dagger - i g\tilde H^\dagger W_\mu - i\frac{g'}{2}Y_\mu \tilde H^\dagger.
\qquad
\end{eqnarray}
Here we have used a matrix notation such as
\begin{eqnarray}
W_\mu = W_\mu^a T^a,\quad 
\Phi = \Phi^\alpha T^\alpha,\quad T^a = \frac{\tau^a}{2},\quad 
T^0 = \frac{{\bf 1}_2}{2},\quad \tr\left[T^\alpha T^\beta\right] = \frac{1}{2}\delta^{\alpha\beta}.
\end{eqnarray}
As usual, it will be convenient to introduce the Weinberg angle $\theta_{\rm w}$ by
\begin{eqnarray}
\tan \theta_{\rm w} = \frac{g'}{g},\quad g_z = \sqrt{g^2 + g'{}^2},\quad e = g_z \sin\theta_{\rm w}\cos\theta_{\rm w},
\end{eqnarray}
and $Z,W$ bosons and the photon fields by
\begin{eqnarray}
Z_\mu = \cos \theta_{\rm w} W_\mu^3 - \sin \theta_{\rm w} Y_\mu,\quad
W_\mu^{\pm} = \frac{W_\mu^1 \mp i W^2_\mu}{\sqrt 2},\quad
A_\mu = \sin \theta_{\rm w} W_\mu^3 + \cos \theta_{\rm w}Y_\mu.
\end{eqnarray}
Then the covariant derivative of the Higgs field can be decomposed as
\begin{eqnarray}
{\cal D}_\mu H = \partial_\mu H - i H\left(e Q A_\mu + g_z T_z Z_\mu 
+ gT_+W_\mu^+ + gT_-W_\mu^-\right),
\end{eqnarray}
with
\begin{eqnarray}
Q = \left(
\begin{array}{cc}
1 & 0 \\
0 & 0 
\end{array}
\right),\quad
T_z = \frac{\tau^3}{2}\cos^2\theta_{\rm w} - \frac{{\bf 1}_2}{2}\sin^2\theta_{\rm w},\\
T_+ = \frac{1}{\sqrt{2}}\left(
\begin{array}{cc}
0 & 1 \\
0 & 0
\end{array}
\right),\quad 
T_- = \frac{1}{\sqrt{2}}\left(
\begin{array}{cc}
0 & 0 \\
1 & 0
\end{array}
\right).
\end{eqnarray}

The supersymmetric scalar potential is given by $V = V_D + V_F$ where $D$ and $F$ term potentials are of  the form
\begin{eqnarray}
V_{D,Y} &=& \frac{g'{}^2}{8} \left(HH^\dagger - \tilde H^\dagger \tilde H\right)^2, \\
V_{D,W} &=& \frac{g^2}{4}\tr \left[\left(H^\dagger H - \tilde H \tilde H^\dagger \right)^2\right]
- \frac{g{}^2}{8}\left(HH^\dagger - \tilde H^\dagger \tilde H\right)^2,\\
V_F &=& \beta g^2 \tr \left[\left(\tilde H H - \mu^2 {\bf 1}_2\right)\left(\tilde H H - \mu^2 {\bf 1}_2\right)^\dagger + \Phi^2 \left( H^\dagger H + \tilde H \tilde H^\dagger \right) \right].
\end{eqnarray}
Closely related models (where $g \to 0$ and $\Phi$ becomes a gauge singlet) were intensively studied in \cite{ISS} in the context of dynamical supersymmetry breaking. Following the paper \cite{ISS}, one can conclude that the minimum energy state is given as follows\footnote{Note that for the standard notation of the Higgs VEV, here we use slightly different notation from the one used in \cite{ISS}.}:
\begin{equation}
H=\tilde{H}^\dagger ={(0,  \mu )},\quad \Phi= \begin{pmatrix} 
 x & 0 \\ 
 0 & 0 \\ 
 \end{pmatrix}\label{vacuum}.
\end{equation}
The vacuum has a positive energy $|h\mu^2|^2$, so that supersymmetry is spontaneously broken.

At classical level, $x$ is a pseudo moduli which parametrizes classical vacua while it gets quantum corrections from one-loop contribution. In this model, the one-loop Coleman-Weinberg potential lifts the pseudo moduli  and stabilizes it to the origin. Since our motivation is mainly to understand classical behaviors of the Nambu monopole and the electroweak string, we treat the pseudo moduli as a parameter of the vacua. 
As we will show in the Appendix, one can stabilize it away from origin either by turning on superpotential of higher order or by adding a higher dimensional K\"ahler potential. However, to keep the model simple, here we do not add such terms and treat the moduli as a parameter. 

On the vacuum \eqref{vacuum}, the electroweak symmetry $U(1)_{\rm Y} \times SU(2)_{\rm W}$ is broken down to $U(1)_{\rm em}$. The masses of the weak bosons are given by
\begin{eqnarray}
m_Z = g_z\mu,\quad
m_W = g\sqrt{\mu^2+\frac{x^2}{2}}.\label{eq:ZWmass}
\end{eqnarray}
The mass of the $W$-bosons are different from the standard one because of the pseudo-moduli $x$.
The masses of the scalar fields can be read from the quadratic terms of small fluctuations
\begin{eqnarray}
\begin{array}{l}
H = \left(a_1+ib_1,\ \mu+a_2+ib_2\right)\\
\tilde H^T = \left(\tilde a_1 + i \tilde b_1,\ \mu + \tilde a_2 + i \tilde b_2\right)
\end{array}
,\quad
\Phi = \left(
\begin{array}{cc}
x+\phi_1 & \phi_2+i\phi_3\\
\phi_2-i\phi_3 & \phi_4
\end{array}
\right).
\end{eqnarray}
The quadratic terms in each potential are given by
\begin{eqnarray}
V_{D,Y}^{(2)} &=&  \frac{\mu^2 g'{}^2}{2} (a_2-\tilde a_2)^2,\\
V_{D,W}^{(2)} &=& \frac{\mu^2 g^2}{2} \left((a_1-\tilde a_1)^2+ (a_2- \tilde a_2)^2+(b_1+\tilde b_1)^2 \right),\\
V_{F}^{(2)} &=& \beta g^2 \bigg( \mu^2\left((a_1-\tilde a_1)^2+(a_2+\tilde a_2)^2+(b_1+\tilde b_1)^2+(b_2+\tilde b_2)^2
+2  \phi_4^2 \right)
\nonumber\\
&+& (a_1 x+ \mu  \phi_2)^2+(\tilde a_1 x+ \mu  \phi_2)^2+(b_1 x- \mu  \phi_3)^2+(\tilde b_1 x+ \mu  \phi_3)^2\bigg).
\end{eqnarray}

The quadratic terms of the small fluctuations also reduce to a simple formula. 
With these reduced formula of the quadratic terms, one can easily see the masses;
\begin{eqnarray}
M_{b_1-\tilde b_1} = 0,\quad 
M_{a_2+\tilde{a}_2}=M_{b_2+\tilde{b}_2}=\sqrt{\beta}g\mu,\quad M_{a_2-\tilde{a}_2}=\frac{g_z\mu}{ \sqrt{2}},\quad
M_{\phi_4} = 2\sqrt{\beta}g\mu.
\end{eqnarray}
Also, $a_1$, $\tilde{a}_1$ and $\phi_2$ mix to give the following eigenvalues. Analogously, $b_1$, $\tilde{b}_1$ and $\phi_3$ mix and give the same three eigenvalues,
\begin{equation}
M_1=0,\quad M_2=g\sqrt{\beta(x^2+2\mu^2)},\quad M_3=g\sqrt{\beta x^2+(1+2\beta^2)\mu^2}.
\end{equation}
The three massless fields are eaten by the gauge bosons, so that
$Z$- and $W$-bosons get the non-zero masses.
On the other hand $\phi_1$ remains the physical massless field associated with the flat direction
of $\phi$, which will get a mass from either Coleman-Weinberg potential or non-canonical K\"ahler potential.

Note that $M_2$ and $M_3$ become infinitely heavy 
for $x \to \infty$ limit (The $W$-bosons are also infinitely heavy and disappear from
the model).
Furthermore, when $\mu=0$, the Higgs field $H$ does not develop any VEV, so that only $a_1+ib_1$ is
the massive field while all other fields  are massless. The symmetry breaking is $U(1)_{\rm Y} \times SU(2)_{\rm W} \to U(1)_{\rm em} \times U(1)_Z$
in this case.

There exists a rich set of  (non-)topological excitations in our model.
They are classical solutions of the equations of motion.
The equations of motion for $H$ are given by
\begin{eqnarray}
{\cal D}^\mu {\cal D}_\mu H + \frac{g'{}^2-g^2}{4}\left(HH^\dagger - \tilde H^\dagger \tilde H\right)H + 
\frac{g^2}{2}H\left(H^\dagger H - \tilde H \tilde H^\dagger\right) \nonumber\\
+\beta g^2 \left(\tilde H^\dagger \left(\tilde H H - \mu^2 {\bf 1}_2\right) + H \Phi^2\right) = 0.
\end{eqnarray}
Exchanging $H$ and $\tilde H^\dagger$ in this equation, we get the equations of motion for $\tilde H^\dagger$.
The equations of motion for $\Phi$ are given by
\begin{eqnarray}
{\cal D}^\mu{\cal D}_\mu \Phi + \frac{\beta g^2}{2}\left\{\Phi,H^\dagger H + \tilde H \tilde H^\dagger\right\} = 0.
\end{eqnarray}
Also we have
\begin{eqnarray}
{\cal D}_\nu W^{\mu\nu} &=& \frac{ig}{2}\left( H^\dagger {\cal D}^\mu H - {\cal D}^\mu H^\dagger H + \tilde H {\cal D}^\mu\tilde H^\dagger - {\cal D}^\mu \tilde H \tilde H^\dagger \right) + i g \left[\Phi,{\cal D}^\mu \Phi\right]\nonumber\\
&-& \tr\left[\frac{ig}{2}\left( H^\dagger {\cal D}^\mu H - {\cal D}^\mu H^\dagger H + \tilde H {\cal D}^\mu\tilde H^\dagger - {\cal D}^\mu \tilde H \tilde H^\dagger \right) + i g \left[\Phi,{\cal D}^\mu \Phi\right]\right] {\bf 1}_2,\\
\partial_\nu Y^{\mu\nu} &=& \frac{ig'}{2}\left({\cal D}^\mu H H^\dagger - H {\cal D}^\mu H^\dagger
+ {\cal D}^\mu \tilde H^\dagger \tilde H - \tilde H^\dagger {\cal D}^\mu \tilde H\right).
\end{eqnarray}

\subsection{The $Z$-string}

In order to get the so-called $Z$-string solution, we make the following ansatz
\begin{eqnarray}
H = \tilde H^\dagger = \left(0,\ h(x_1,x_2)\right),\quad
\Phi = \left(
\begin{array}{cc}
0 & 0 \\
0 & 0
\end{array}
\right),\quad W_\mu^{\pm} = A_\mu = 0,\quad Z_\mu = Z_\mu (x_1,x_2).
\end{eqnarray}
Then the above equations of motion reduce to the following simple equations
\begin{eqnarray}
{\cal D}_i{\cal D}^i h + \beta g^2 h\left(|h|^2 - \mu^2\right) = 0,\label{eq:ano1}\\
\partial_j Z^{ij} = - i g_z\left(\bar h {\cal D}^i h - {\cal D}^i\bar h h\right),
\label{eq:ano2}
\end{eqnarray}
with $i=1,2$ and 
\begin{eqnarray}
Z_{\mu\nu} = \partial_\mu Z_\nu - \partial_\nu Z_\mu,\quad {\cal D}_\mu h = \partial_\mu h + \frac{i g_z}{2} Z_\mu h.
\end{eqnarray}
The Eqs.~(\ref{eq:ano1}) and (\ref{eq:ano2})  are the well-known differential equations of the ANO vortex in the Abelian-Higgs model \cite{Abrikosov:1956sx}.

Although Eqs.~(\ref{eq:ano1}) and (\ref{eq:ano2}) are the same differential equations for the topologically stable ANO vortex,
the $Z$-string is not supported by any topological stability. This is because the first homotopy group of
the vacuum manifold $[SU(2)_{\rm W} \times U(1)_{\rm Y}] / U(1)_{\rm em} \simeq 
SU(2) \simeq S^3$ of 
the electroweak symmetry breaking is trivial.
To treat such non-topological solitons, it is instructive to start with global limit of $SU(2)_{\rm W}$ group. The limit is the so-called semi-local model.
In this limit, symmetry breaking in the vacuum \eqref{vacuum} involves global and local symmetries and can generate a non-dissipative string. In \cite{EHT}, a semi-local string is constructed in a model closely related to our model. 
As is well known, local stability of the Abelian semi-local string depends on the coupling constants of the model. In our model, they are $g$ and
$g_z$. When $g < g_z$, the string is called type I, and the semi-local string is stable. On the other hand, when $g > g_z$, the string
is called type II and it is unstable to be diluted. The semi-local string is also marginally stable at the critical point $g = g_z$ which is called BPS.
It is also easy to see global stability of the string by looking at the kinetic term in Lagrangian. For the finite energy string, the  field  configuration at spatial infinity has to be in gauge orbit. In this configuration, a term in the energy density, $\mathcal{E}\sim |D_i H|^2$ goes like $1/r$. 
In the global limit of the $SU(2)$ group, only $U(1)$ orbits are gauge orbits inside the vacuum manifold.
Deviation from the $U(1)$ orbits increases the energy coming from the kinetic terms, 
and consequently semi-local vortices are stable.

On the other hand, when the  $SU(2)$ group is gauged, by using the $SU(2)$ orbit away from the $U(1)$ orbit, 
one can decrease the energy continuously with finite energy cost. 
This implies that the string becomes both locally and globally unstable and decay. 
The instability can be shown explicitly by analyzing small fluctuations of bosonic fields around the string solution in our model.
This is rather complicated analysis but
ignoring the fluctuation of $\delta \Phi$, then it is simply reduced to the one for the standard electroweak string \cite{JamesI}. 
Since unstable modes already exist in the fluctuation of the Higgs fields, 
the additional field $\Phi$ does not cure the stability of the string at $x=0$.

On the other hand, when $x\to \infty$ one can show that the string becomes stable. To see that, consider first integrating out of the infinitely massive field with the mass scale $x$. As we have seen, the first components of the Higgs fields and the $W$-bosons are integrated out. So we are left with Lagrangian of the $U(1)_{\rm Y} \times U(1)_{\rm W}$ gauge field with the second Higgs field and $\Phi$. When the second Higgs field develops a non-zero
VEV, $U(1)_{\rm Y} \times U(1)_{\rm W}$ breaks to $U(1)_{\rm em}$. The string generated with this symmetry breaking is the ANO type string which is stable against  fluctuations. With these facts one can conclude that by changing the pseudo moduli $x$ one can interpolate from a electroweak string to ANO one.

\subsection{Nambu monopoles}

As pointed out by Nambu \cite{Nambu}, the electroweak string can end on a non-isolated monopole, which is called the Nambu monopole. By spontaneous nucleation of the monopole/aniti-monopole pair, the $Z$-string can decay. Since one of our motivation in this paper is to study the monopole and string composite, here we would like to summarize basics of such a composite. 

The Nambu monopole terminates the magnetic $Z$-flux coming from the $Z$-string. Also it carries the magnetic monopole charge for the electromagnetic $U(1)$ field,  
\begin{equation}
F_A = {{4\pi} \over {g_z}} \tan\theta_{\rm w}
                  = {{4\pi} \over {e}} \sin^2 \theta_{\rm w} 
\end{equation}
To estimate the mass of the Nambu monopole, let us focus on the $D$-flat direction $H=\widetilde{H}^{\dagger}$ and take the bosonic part of $\Phi$ to be zero. On this point, the potential term becomes similar to the one discussed in \cite{Nambu}
\begin{eqnarray}
V_{\rm red} = \beta g^2 \tr\left[ \left( H^\dagger H - \mu^2 {\bf 1}_2\right)^2 
\right]. \label{redpot}
\end{eqnarray}
Therefore following the argument shown in \cite{Nambu}, one can obtain the mass $M_{\rm Nambu}$ and the radius $R$ of the Nambu monopole as follows: 
\begin{equation}
M_{\rm Nambu}= {4 \pi \over 3 e} \sin ^{5/2}\theta_{\rm w} \sqrt{m_H \over m_W}  \mu ,\quad R = \sqrt{ \sin\theta_{\rm w}  \over m_H m_W},
\end{equation}
where $m_W=g\mu$ and $m_H=\sqrt{\beta}g \mu$. With the precise LHC date for Higss boson $m_H\simeq 125$GeV and other dates, $\sin^2  \theta_{\rm w}  \simeq 0.231$, $m_W\simeq 80.3$GeV, $m_Z\simeq 91.1$GeV, we can precisely estimate the mass of the Nambu monopole,
\begin{equation}
M_{\rm Nambu}\simeq 689 \ {\rm GeV}. \label{NambuMass}
\end{equation}

\begin{figure}[htbp]
\begin{center}
\includegraphics[height=4cm]{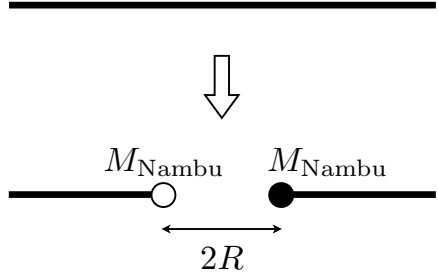}
\end{center}
 \vspace{-.5cm}
\caption{\sl \small A long string decays with a monopole/anti monopole pair creation. 
\label{fig:string_break}
}
\end{figure}

Finally we quickly review decay probability of the $Z$-string by following \cite{Preskill}. Since the string can break due to the nucleation of a monopole/antimonopole pair, there is a nonzero probability of decay. Suppose that there is a long straight string with tension $T_{Z}$. It  will tunnel to a configuration in which a monopole/anti monopole pair is created with separation $2R$. The energy cost of producing the pair is $2M_{\rm Nambu}$,
see Fig.~\ref{fig:string_break}. This cost must be balanced by 
the vortex energy  $2R\,T_{Z}$    between the monopole pair, when the monopole pair is created. Thus, the initial separation of the monopole and anti-monopole is $R={2M_{\rm Nambu}/ T_{Z}}$. 

The bounce action of this process is given by
\begin{equation}
B= 2\pi RM_{\rm Nambu} - \pi R^2T_{Z}.
\end{equation}
and the decay probability is roughly given by $P \propto
\exp(-B)$. 

\section{Geometric Realization}\label{sec:geometric}

\subsection{D-brane set up}

In this section, we will show a geometric interpretation of the Nambu monopole and the $Z$-string in the context of Type IIA string theory. A virtue of promoting the Higgs sector to a supersymmetric one is a simple realization in string theory which visualizes various aspects of monopole, string and their composites in a geometrical way. In the papers \cite{BraneI,BraneII,BraneIII,Kutasov1}, the original Intriligator-Seiberg-Shih (ISS) model is embedded in Type IIA string theory and the metastable state of the model is described by a geometric configuration of intersecting branes. Here, following \cite{BraneI,BraneII,BraneIII,Kutasov1} (see \cite{KOOreview} for a review), we construct a brane configuration of our model with 
$SU(2) \times U(1)$ gauge group as follows, see Figure \ref{fig:BraneI}(a):
\begin{figure}[htbp]
\begin{center}
  \epsfxsize=7cm 
\begin{tabular}{cc}
\epsfbox{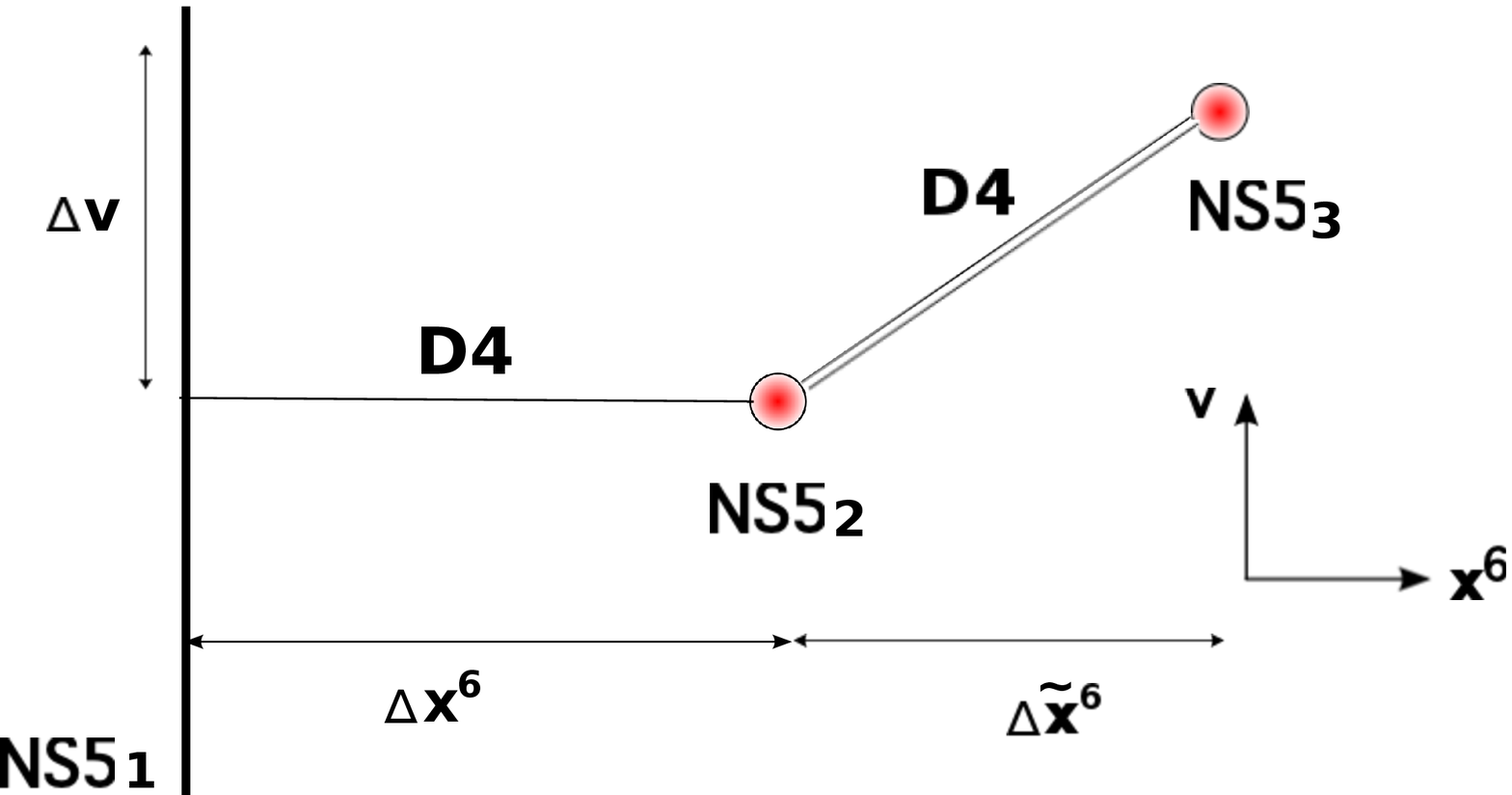} &
\epsfbox{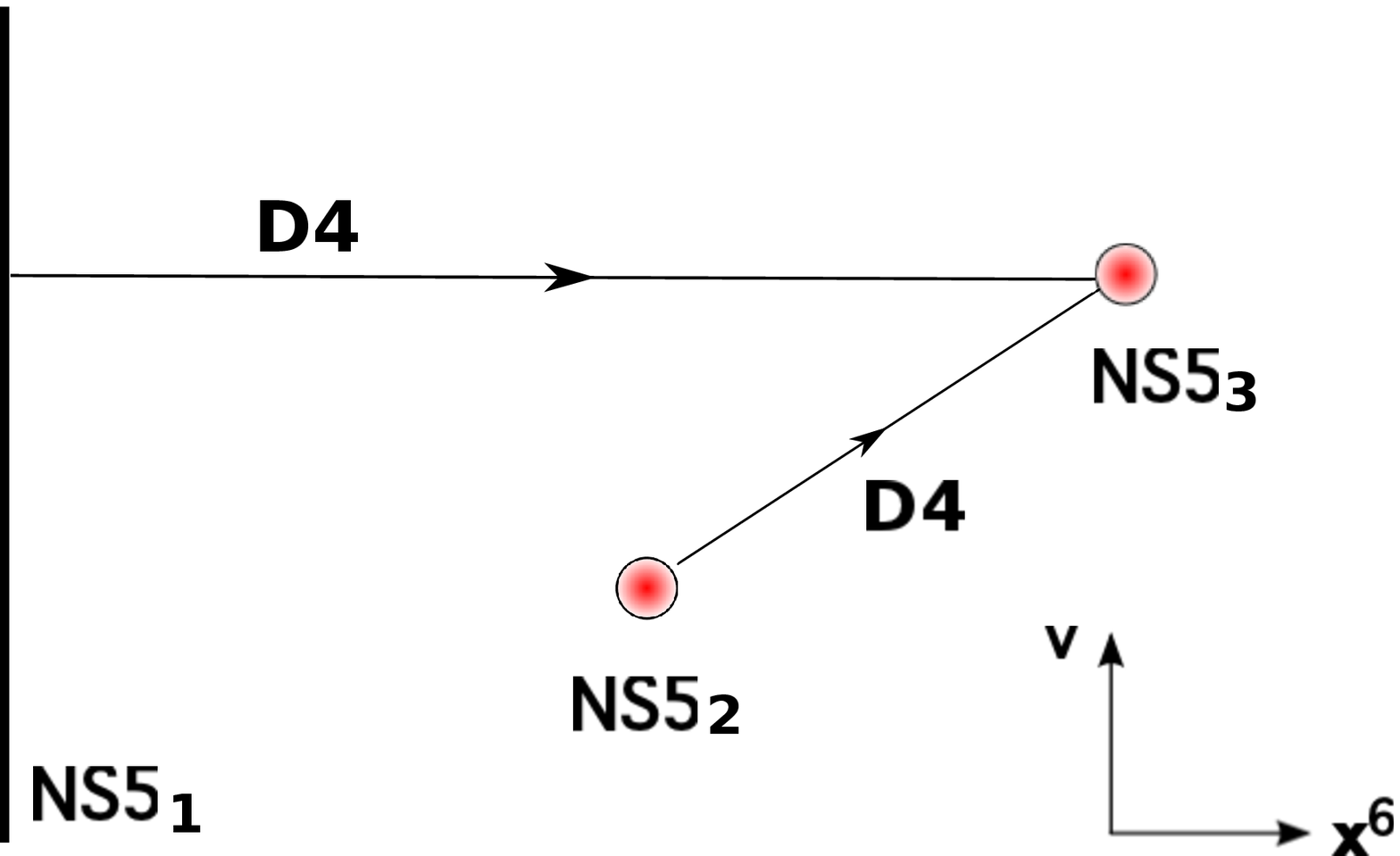}\\
(a) & (b)
\end{tabular}
\end{center}
 \vspace{-.5cm}
\caption{\sl \small Brane configuration of the standard model 
before (a) and after (b) the Higgs mechanism.  
(a) The gauge group is $SU(2)_{\rm W} \times U(1)_{\rm Y}$ where 
$SU(2)_{\rm W}$ and $U(1)_{\rm Y}$ are associated with the two tilted D4-branes and 
one horizontal D4-brane, respectively. 
(b) After the reconnection of the two D2 branes, the symmetry breaking occurs 
as  $SU(2)_{\rm W} \times U(1)_{\rm Y} \to U(1)_{\rm em}$, 
where one remaining tilted D4-brane corresponds to $U(1)_{\rm em}$. 
\label{fig:BraneI}
}
\end{figure}

\begin{itemize}
\item One NS5 brane whose world-volume extends to the (0123) and (45) directions and
located at $w, x^6, x^9=0$, with $w \equiv x^7+ix^8$. 
We call this as the NS5$_1$ brane. 

\item One NS5 brane whose world-volume extends to the (0123) and (78) directions and
located at $v, x^9=0$ and $x^6=\Delta x^6$, where
$v \equiv x^4+ix^5$ and $\Delta x^6 >0$. 
We call this as the NS5$_2$ brane. 
 
\item One NS5 brane whose world-volume extends to 
the (0123) and (78) directions and
located at $v= \Delta v$ and
$x^6 = \Delta x^6 +\widetilde{\Delta x^6}$.  We call this
as the NS5$_3$ brane. 

\item One D4 brane stretched in the (0123) directions and 
going between the NS5$_1$ and NS5$_2$ branes along the $x^6$ axis. They
are located at $v, w, x^9=0$. 

\item Two D4 branes whose world-volumes extend to the (0123) directions and stretched between the NS5$_2$ brane at $x^6=\Delta x^6$ and the NS5$_3$ brane at $x^6=\Delta x^6+\widetilde{\Delta x^6}$. 
They are tilted in the (45) and 6 spaces. 
\end{itemize}

Various field theory parameters in our model are identified with parameters in string theory as follows: 
\begin{equation}
g_{U(1)}^2={ g_sl_s\over \Delta x^6 }\ , \qquad g_{SU(2)}^2={g_sl_s\over \widetilde{\Delta x^6}}\ ,
\qquad \mu^2=-{\Delta v\over  g_s l_s^3}, \qquad m={\Delta w \over   l_s^2}. \label{parameters}
\end{equation}
Here to avoid confusion, we explicitly write gauge couplings for $U(1)_{\rm Y}$ and $SU(2)_{\rm W}$ as $g_{U(1)}$ and $g_{SU(2)}$ rather than $g$ and $g^{\prime}$.

To realize field theory as a low-energy effective theory on D-branes, we first take the so-called brane limit in which $g_s$ is infinitesimal and all the length scales of the system are larger than $l_s$. In this limit, the degrees of freedom are open strings and the dynamics is described by the Dirac-Born-Infeld action of a set of D-branes. In this low-energy theory, we have infinite towers of massive modes with masses of order ${\cal O}((\Delta x^6)^{-1})$ or ${\cal O}((\widetilde{\Delta x^6})^{-1})$ coming from the Kaluza-Klein reduction of open strings. To make these heavy, we should send $\Delta x^6$, $\widetilde{\Delta x^6}$, $l_s$ and $\Delta v$ to zero while keeping their ratios fixed, as in Eq.~\eqref{parameters}. This limit is called the decoupling limit. As we will comment below, there is a subtlety in taking the limit when we identify the field theory.

\subsection{Vortices and Monopoles via D-branes}

Now, we are ready to study vacuum structure of our theory in the brane setup. Open strings stretched  between the single D$4$-brane and the two tilted D$4$-branes correspond to the Higgs fields $H, \tilde{H}$ in the field theory. 
Clearly, the brane configuration in Fig.~\ref{fig:BraneI}(a) is unstable. 
The reconnection of the two D4 branes reduces energy of the configuration 
as in  Fig.~\ref{fig:BraneI}(b).
This process can be understood as giving VEV to the Higgs field and the symmetry breaking $U(1)\times SU(2)\to U(1)_{\rm em}$ happens. At the classical level, the remaining tilted single D4 brane can slide freely in the $x^7$-$x^8$ directions, 
whose degrees of freedom corresponds to the pseudo-moduli in the vacuum. 
This aspect nicely reproduces the field theory results. 

\begin{figure}[htbp]
\begin{center}
  \epsfxsize=7cm 
\begin{tabular}{cc}
\epsfbox{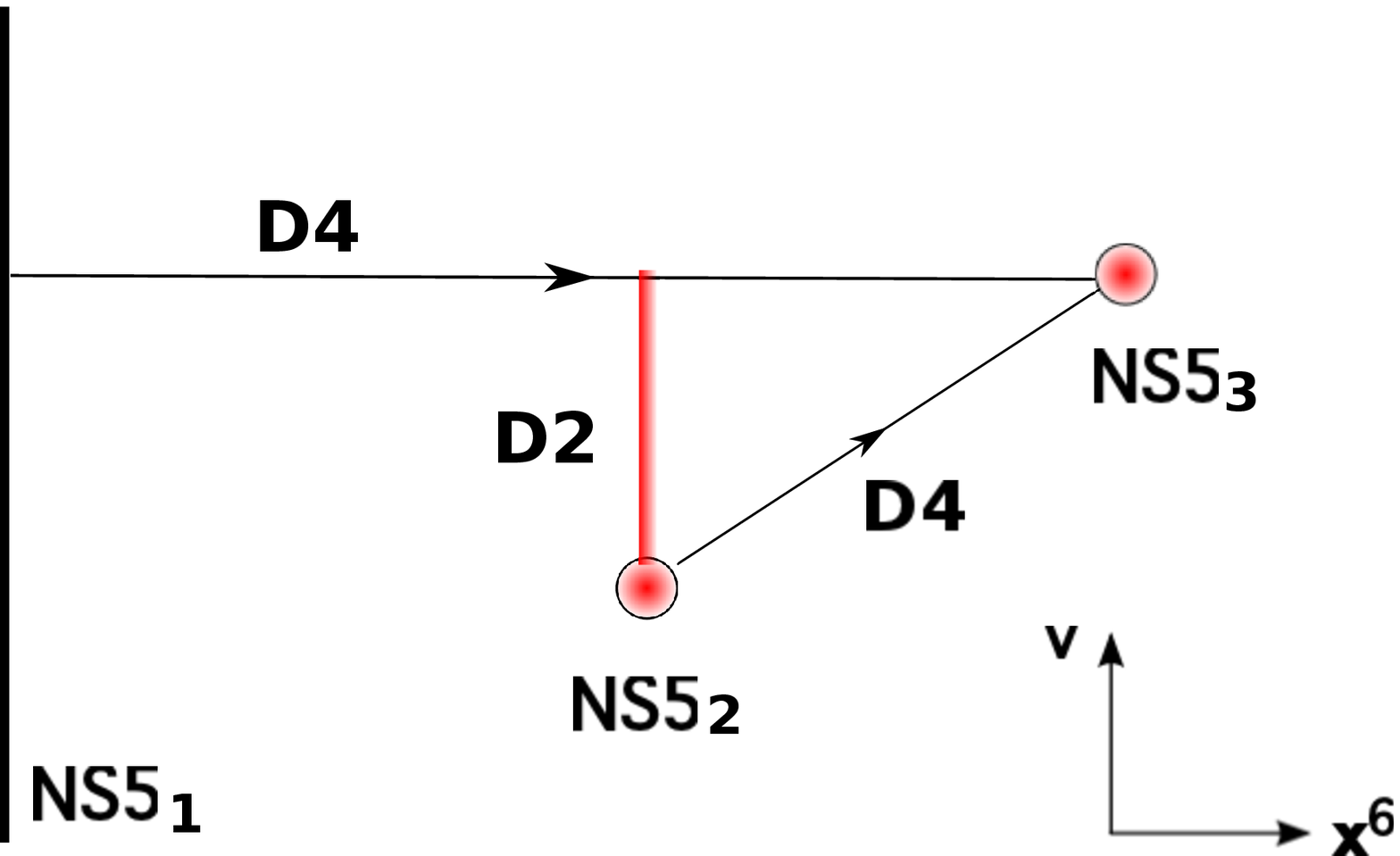} &
\epsfbox{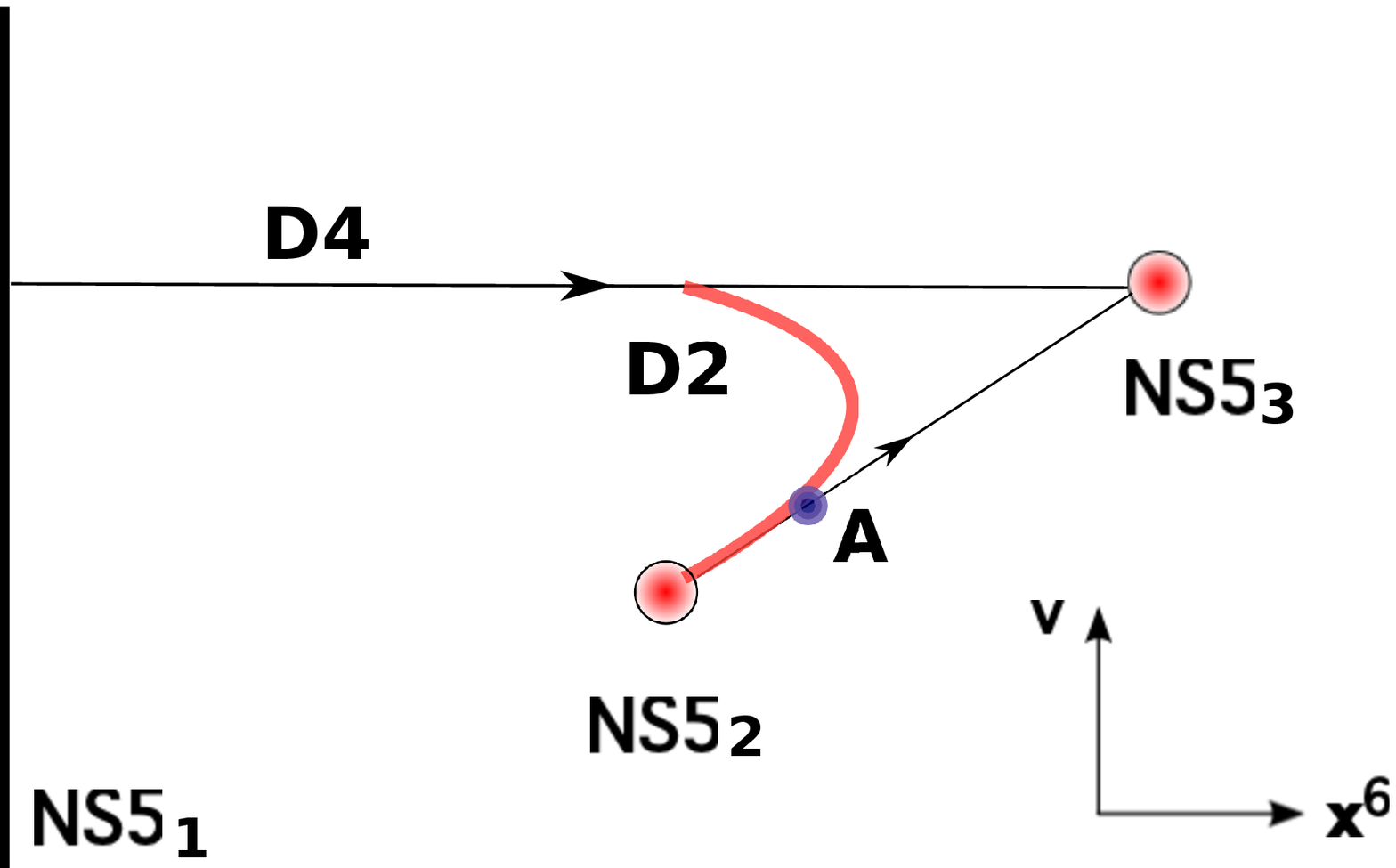}\\
(a) & (b)
\end{tabular}
\end{center}
\caption{\sl \small Brane configurations for  (a) a $Z$-string and (b) its decay.  \label{fig:BraneII}}
\end{figure}

In this brane setup, one can identify the $Z$-string as a D2 brane by following the argument 
in \cite{EHT}, in which a semi-local vortex present in the magnetic description of the ISS model was studied. 
Our model is closely related to that model, so it is natural to consider the D2-brane connecting 
the D4-brane and the NS5$_2$-brane along the $v$ direction as seen in 
Fig.~\ref{fig:BraneII}(a). 
Since this string is generated under the symmetry breaking of the $U(1)_Z$, we interpret it as the $Z$-string. As emphasized in the previous section, when $SU(2)$ symmetry is gauged, the $Z$-string is unstable. As pointed out in \cite{IbeOokouchi}, it is interesting to see this process in the context of the brane picture. One of the key aspects of the semi-local vortex is the existence of open strings stretched between the tilted D4-brane and the D2-brane. Because of the additional degrees of freedom, the semi-local vortex has the size and phase moduli. When gauging the global symmetry, what happens to this degree of freedom? Actually, it develops a tachyonic mode and makes the D2 brane unstable. To see that, let us consider deformation of the D2 brane, and push it toward the $x^6$-direction as shown in Fig.~\ref{fig:BraneII}(b).  From the point A to NS5$_2$, the D2 brane is parallel to the tilted D4 brane. In this case, the open string connecting these two branes includes a tachyonic mode. It is easy to see such the tachyonic mode by using T-duality: Taking T-dual twice, this system becomes intersecting D4-branes\footnote{For example, if the vortex is  extending along the $x^1$-direction, we can take T-duals in the $x^2$ and $x^9$ directions.}. After the tachyon condensation occurs, the segment of the D2-brane between the point A and NS5$_2$ is smeared out and eventually disappears. Finally the point A moves toward the NS5$_3$ and the whole D2-brane disappears by the same process. Since the D2-brane carries the magnetic flux of the broken $U(1)_Z$ symmetry, this process can be interpreted as dissipative process of the $Z$-flux.

One interesting situation occurs when the VEV of the pseudo moduli is nonzero, $x\neq 0$. In this case, the tachyonic mode in the D2-D4 string does not exist. Since this mode is massive in low energy, it can be integrated out. The D2-brane configuration in such situation is locally similar to the Hanany-Tong's D-brane configuration \cite{Tong,TongII} for non-Abelian vortices \cite{Auzzi:2003fs,Eto:2005yh}, thus we can interpret the D2-brane as an ANO string, which nicely reproduces the field theory arguments. However, this string is just merely metastable in a local minimum of the potential. By quantum mechanical tunneling or thermal fluctuations, the D2-brane can reach the tilted D4 brane and decays through the same process described above.

\subsection{Monopole-Vortex complex}

The main ingredient of our study is a monopole-vortex complex. 
A BPS monopole-vortex composite was realized 
by a single D2 brane in ${\cal N}=2$ supersymmetric gauge theories 
\cite{Hanany:2004ea,Auzzi:2004yg}. 
In this section, we show that our non-BPS monopole-vortex composite 
can also be realized by a single D2-brane. 

\begin{figure}[htbp]
\begin{center}
  \epsfxsize=7cm 
\begin{tabular}{cc}
\epsfbox{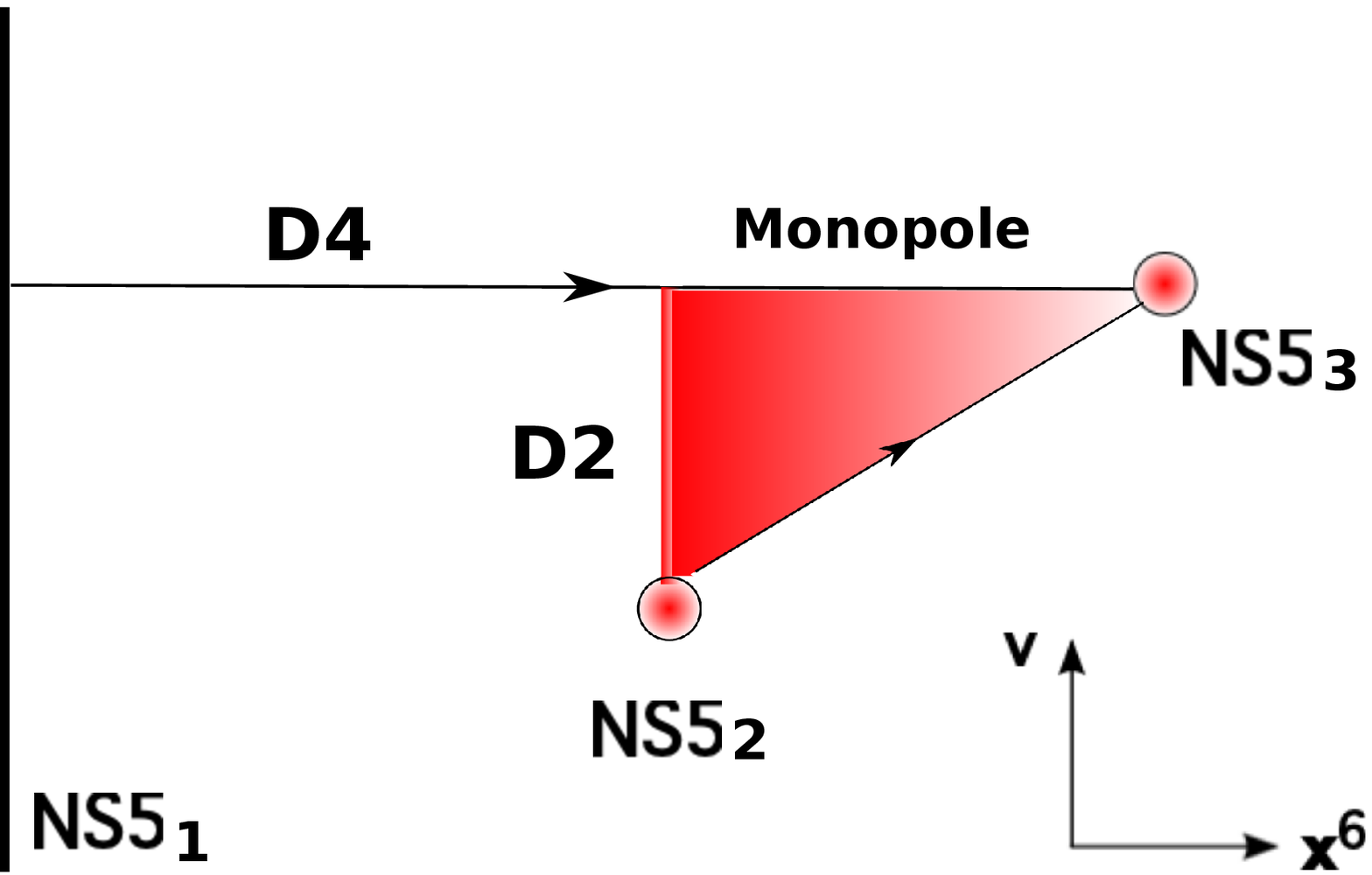} &
\epsfbox{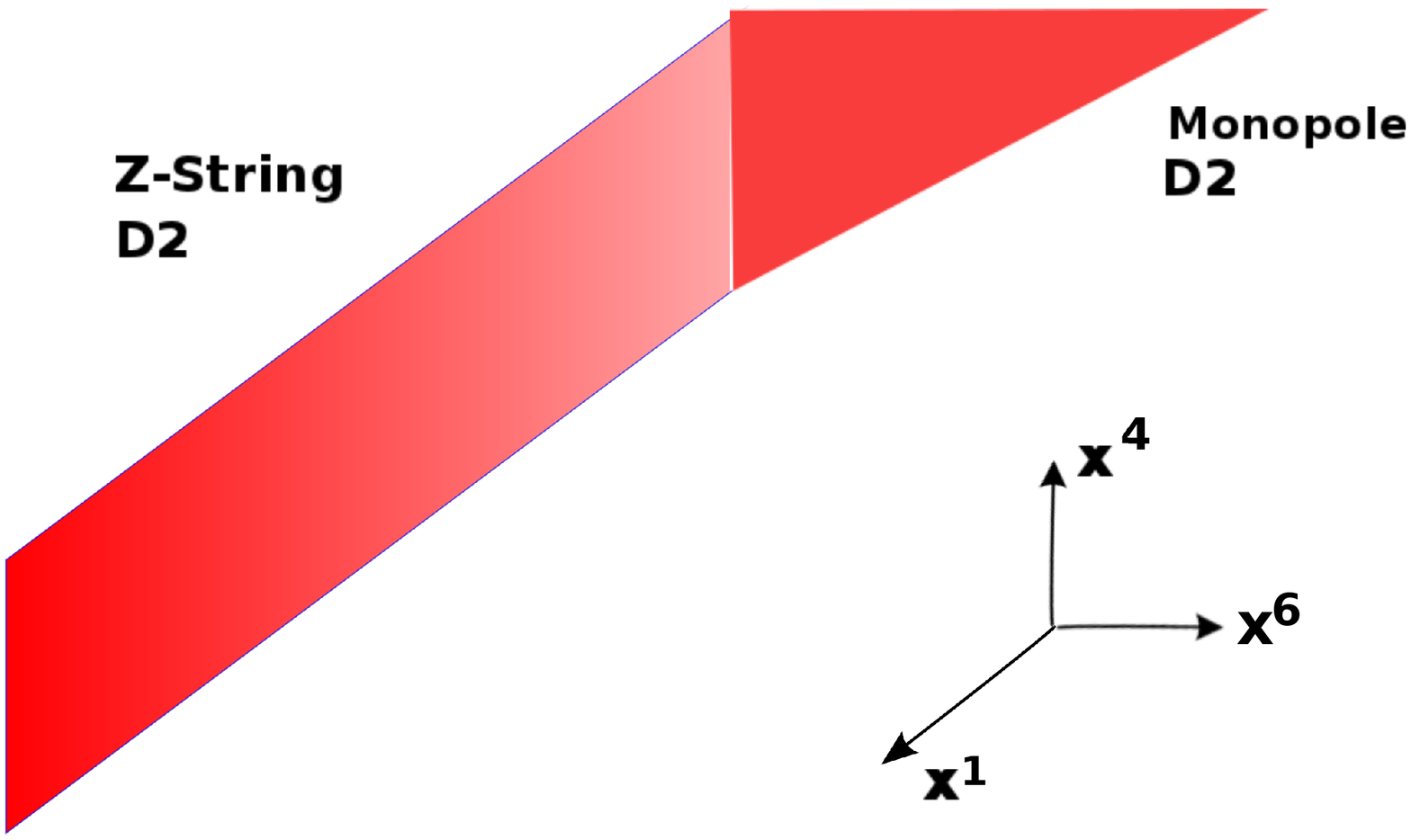}\\
(a) & (b)
\end{tabular}
\end{center}
\caption{\sl \small Brane configurations for a Nambu monopole and a $Z$-string ending on it.  \label{fig:BraneIII}}
\end{figure}

Here again, we consider two cases $x=0$ and $x\neq 0$ separately. When $x=0$, as was pointed out by Nambu, the electroweak string can end on the Nambu monopole. First step is to identify the Nambu monopole in the context of the brane configuration. Since the Nambu monopole is not isolated, it takes an infinite amount of energy to create a single monopole. On the other hand, in the presence of the electroweak string, a pair of a monopole and an anti-monopole can be produced with finite energy. Therefore, in a geometric picture, such a monopole should exist only when the D2-brane corresponding to the vortex exists. A D2-brane surface surrounded by the D2-brane corresponding to the vortex and the two D4-branes, 
shown in the triangle region in Fig.~\ref{fig:BraneIII}(a), is a natural candidate for the Nambu monopole. 
The D2-brane corresponding to the vortex is necessary for the existence of the Nambu monopole,  
and actually that vortex D2-brane is extended along the $x^1$-direction.
One reaches a single semi-infinite D2-brane bent into the internal space at $x^1=0$, 
as shown in Fig.~\ref{fig:BraneIII}(b). 
This explains geometrically why the Nambu monopole cannot exist alone and it requires a $Z$-string in field theory. Moreover, the triangle region of the D2-brane corresponding to the monopole ends on the two D4-branes with codimension two on the D4-brane world-volumes. As is known in the D-brane configuration  for a `t Hooft-Polyakov monopole \cite{GiveonKutasov}, this D2-brane carries the magnetic charge for the unbroken gauge group. Thus, in our situation, this D2-brane corresponding to the monopole carries the $U(1)_{\rm em}$ magnetic charge, which is also precisely the case of the Nambu monopole. 
In the brane picture, both the Nambu monopole and the $Z$-string are expressed by a single D2-brane 
at the same time. 

\begin{figure}[htbp]
\begin{center}
  \epsfxsize=8cm \epsfbox{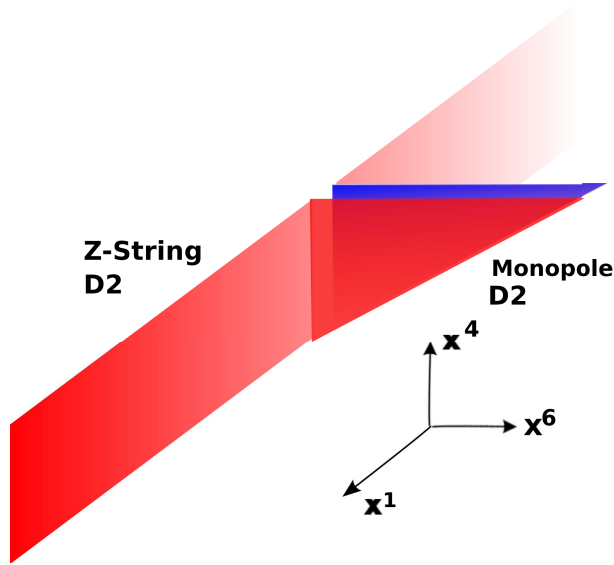}
\end{center}
 \vspace{-.5cm}
\caption{\sl \small Brane configuration for decaying process of a $Z$-string 
by creating a pair of a Nambu monopole and an anti-Nambu monopole.  \label{fig:BraneIV}}
\end{figure}

Next, let us discuss the decay of a $Z$-string by creation of a pair of 
a Nambu monopole and an anti-monopole.
Suppose an infinite D2-brane corresponding to the vortex is extended along the $x^1$-direction.
This $Z$-string can decay if it is bent into the internal space at $x^1=0$. 
The bent D2-brane ends on the two D4-branes, 
as shown in Fig.~\ref{fig:BraneIV}. 
One can interpret the D2-branes in the internal space as the Nambu monopole and anti-Nambu monopole. Thus, spontaneous nucleation of a pair of the monopole and anti-monopole breaks the $Z$-string. This perfectly matches with the field theory argument. With these successes, we identify the D2-brane wrapping on the triangle as the Nambu monopole. 

To see more precise correspondence with the brane configuration, let us consider monopole mass. Using the D2 brane tension, we can estimate the mass of the monopole as follows:
\begin{equation}
M_{\rm Nambu}\sim T_{D2} \Delta v \widetilde{\Delta x^6}
= \left({\Delta v \over g_s l_s^3}\right) \left({ \widetilde{\Delta x^6} \over g_s l_s}\right) g_s l_s 
={g_s l_s \over g_{SU(2)}^2} \mu^2.
\end{equation}
At first sight, this is inconsistent with the field theory, because in the field theory limit argued before gives us zero monopole mass \eqref{NambuMass}. We claim that this discrepancy comes from the massive gauge boson in the brane description. In the brane setup, after the Higgsing some of gauge fields are gone by the boundary condition. Thus, the corresponding massive gauge boson should be identified with the Kaluza-Klein (KK) modes in the $x^6$-direction. However, under the field theory limit, it becomes infinitely massive. So it is reasonable to conclude that the low-energy effective theory on the D-brane may correspond to the one for infinitely massive gauge boson. This interpretation agrees with the mass formula in the field theory. By taking $m_W\to \infty$ limit in the mass formula, one sees that mass of the monopole vanishes.

With this understanding, let us take a non-standard limit of the theory to produce with a finite W-boson mass case.  Since we claimed that a KK mode in the $x^6$-direction corresponds to the massive W-boson, if we take $(\widetilde{\Delta x^6} )^{-1} \sim O(K\mu)$, then the W-boson becomes in the same energy scale of the field theory, where $K$ is a dimensionless parameter which will be fixed below. Apparently, this is not the field theory limit but the so-called brane limit, so there should exist various corrections from stringy effects. However, since we are interested in reproducing the classical results of field theory, we ignore such corrections and discuss only classical aspects of the brane configuration. However, it may be interesting that quantum corrections to non-BPS objects can be interpreted as stringy correction effects in our limit.  
In this limit, the Nambu monopole mass is given by 
\begin{equation}
M_{\rm Nambu}\sim T_{D2}\Delta v \widetilde{\Delta x^6}\sim K^{-1} \mu .
\end{equation}
Therefore, by taking dimensionless parameter $K$ appropriately, we can reproduce the correct mass for the monopole, 
\begin{equation}
K^{-1}={{1\over e} \sin^{5/2} \theta_{\rm w} \sqrt{m_H \over m_W}}.
\end{equation}

\subsection{Nambu and `t Hooft-Polyakov monopole complex}

One of advantages of this argument can be seen in the following calculation of the binding energy of the Nambu monopole and `t Hooft-Polyakov monopole. 
The binding energy between a monopole and a vortex 
was studied before in Ref.~\cite{Auzzi:2004yg}. 
In this section, we concentrate on the binding energy between the `t Hooft-Polyakov and the Nambu monopoles ignoring the binding energy 
between the monopole and the electroweak string.

When $x\neq 0$, there exists the `t Hooft-Polyakov monopole associated with the intermediate spontaneous symmetry breaking $SU(2)\to U(1)$, where $U(1)$ is finally broken to nothing by the Higgs VEV. 
The 't Hooft-Polyakov monopole can be best seen in the limit of $\mu \to 0$ in which two D4-branes are parallel.
As is well known, the Hanany-Witten brane realization of the 't Hooft-Polyakov monopole is a D2-brane suspended by the parallel D4-branes and NS5$_{2,3}$. Its mass is proportional to the area of the flat D2-brane, namely, $\sim m/g^2$. 

\begin{figure}[htbp]
\begin{center}
  \epsfxsize=10cm \epsfbox{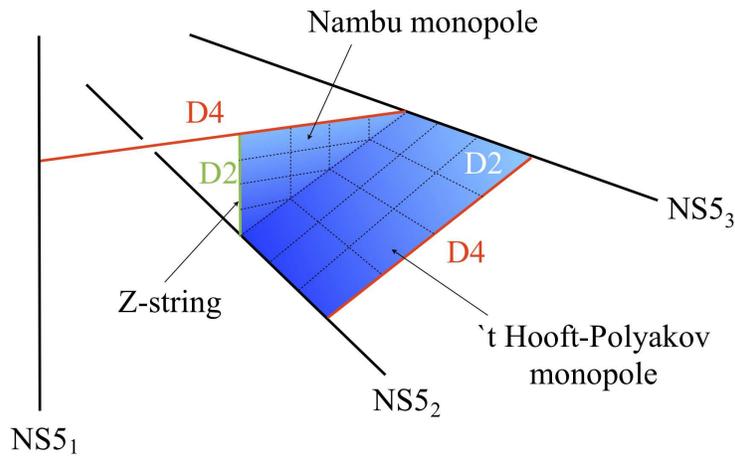}
\end{center}
 \vspace{-.5cm}
\caption{\sl \small Nambu and 't Hooft Polyakov monopoles. }\label{conBrane-tHooft+Nambu}
\end{figure}

\begin{figure}[htbp]
\begin{center}
  \epsfxsize=10cm \epsfbox{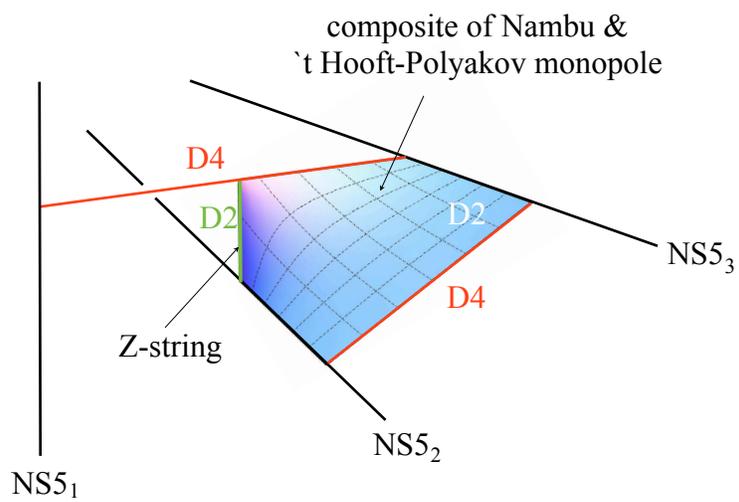}
\end{center}
 \vspace{-.5cm}
\caption{\sl \small Minimal surface, bound state of Nambu and t Hooft Polyakov monopole. }\label{conBrane-composite}
\end{figure}

When $\mu > 0$, the two D4-branes are not parallel but are twisted. This implies that the 't Hooft-Polyakov monopole does not exist alone. As can been seen from Fig.~\ref{conBrane-tHooft+Nambu}, it is always accompanied by the Nambu monopole and the $Z$-string, since the D2-brane must end on a D4-brane or stretch infinitely without ending.
Clearly, the D2-brane stretching between the D4-branes and NS5-branes in Fig.~\ref{conBrane-tHooft+Nambu} is not a minimal surface.  By the ``color-flavor'' locking, the Nambu monopole and `t Hooft-Polyakov monopole constitute a bound state as   shown in Fig.~\ref{conBrane-composite}. 

An interesting question is if we can compute the binding energy of the composite. 
As the rigid 't Hooft-Polyakov monopole, the mass of a pointlike object can be identified with the minimal surface area of the D2-brane in internal spaces as in Fig.~\ref{conBrane-composite}.
Then the binding energy can be found by subtracting the surface area of the D2-brane given in Fig.~\ref{conBrane-composite}
from that in Fig.~\ref{conBrane-tHooft+Nambu}.


\clearpage

\begin{figure}[htbp]
\begin{center}
  \epsfxsize=10cm \epsfbox{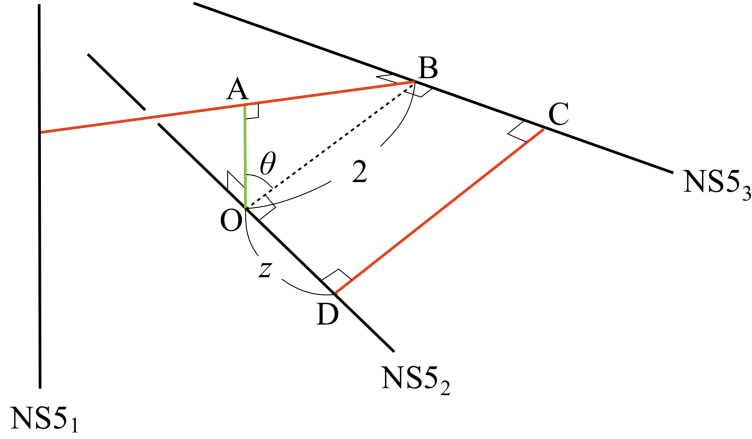}
\end{center}
 \vspace{-.5cm}
\caption{\sl \small Dimensionless expression of the D-brane configuration}\label{dimless}
\end{figure}

Let us show some examples for the calculation below. For simplicity, we will use the following dimensionless
notation where we fix the distance between O and B in Fig.~\ref{dimless} to be 2.  Namely, we have two free
parameters $(\theta,z)$ to identify the D-brane configuration
\begin{eqnarray}
\mu^2 \sim \text{OA} = 2 \cos \theta,\quad
\frac{1}{g^2}\sim \text{AB} = 2 \sin \theta,\quad
m \sim \text{OD} = z.
\end{eqnarray}

\begin{figure}[htbp]
\begin{center}
  \epsfxsize=10cm \epsfbox{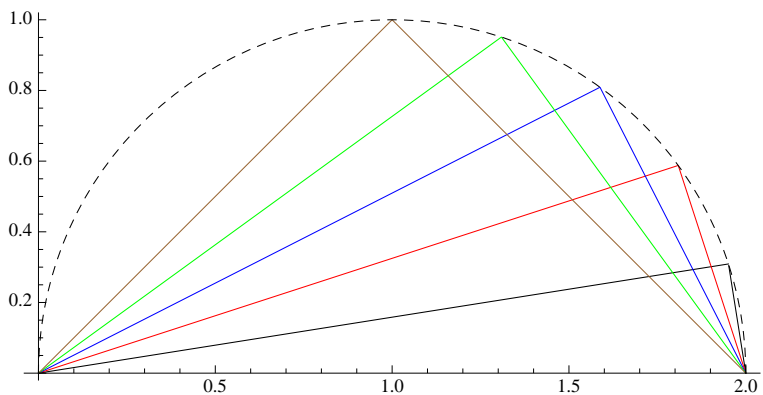}
\end{center}
 \vspace{-.5cm}
\caption{\sl \small Various Dirichlet conditions $f(x)$ with $\theta = \frac{\pi}{20},\ \frac{\pi}{10},\ \frac{3\pi}{20},\ \frac{\pi}{5},\ \frac{\pi}{4}$.}\label{f}
\end{figure}

We can replace the problem of obtaining minimal surface of the D2-brane by an elementary  two-dimensional Laplace equation
with Dirichlet boundary conditions, 
by projecting the surface onto the plane OBCD. 
The boundaries are the segments OB, BC, CD, DO in Fig.~\ref{dimless}.
The height of D2-brane from the plane OBCD is expressed by a function $u(x,y)$. Here $x, y$ are the coordinates on the OBCD plane, 
where we take the $x$-axis along OB and the $y$-axis along OD with the origin O. Then the minimal surface satisfies the Laplace equation
\begin{eqnarray}
\left(\partial_x^2 + \partial_y^2\right) u(x,y) = 0.
\end{eqnarray}
We solve the Laplace equation with the following Dirichlet conditions
\begin{eqnarray}
u(x,0) &=& f(x) = \left\{
\begin{array}{lcc}
\dfrac{\sin 2\theta}{1+\cos 2\theta} x & &{\rm for}\quad 0\le x \le 1+\cos 2\theta\\
&&\\
\dfrac{\sin 2\theta}{1-\cos 2\theta} (2-x) & &{\rm for}\quad 1+\cos 2\theta \le x \le 2
\end{array}
\right.,\\
u(x,z) &=& 0,\\
u(0,y) &=& 0,\\
u(2,y) &=& 0.
\end{eqnarray}
This is analytically solvable. The solution is given by
\begin{eqnarray}
u(x,y) = \sum_{k=1}^\infty g_k \dfrac{\sin{\frac{x}{2}k\pi }\sinh\frac{z-y}{2}k\pi }{\sinh{\frac{z}{2}k\pi}},
\end{eqnarray}
where $g_k$ is the Fourier expansion coefficient of $f(x)$: 
\begin{eqnarray}
f(x) = \sum_{k=1}^\infty g_k \sin \frac{x}{2}k\pi.
\end{eqnarray}
Let us consider five particular cases with $\theta = \frac{\pi}{20},\ \frac{\pi}{10},\ \frac{3\pi}{20},\ \frac{\pi}{5},\ \frac{\pi}{4}$ below.
The configurations as $y=0$ made of the segments AO and OD are given 
in Fig.~\ref{f} for various $\theta$. 

\begin{figure}[htbp]
\begin{center}
  \epsfxsize=16cm \epsfbox{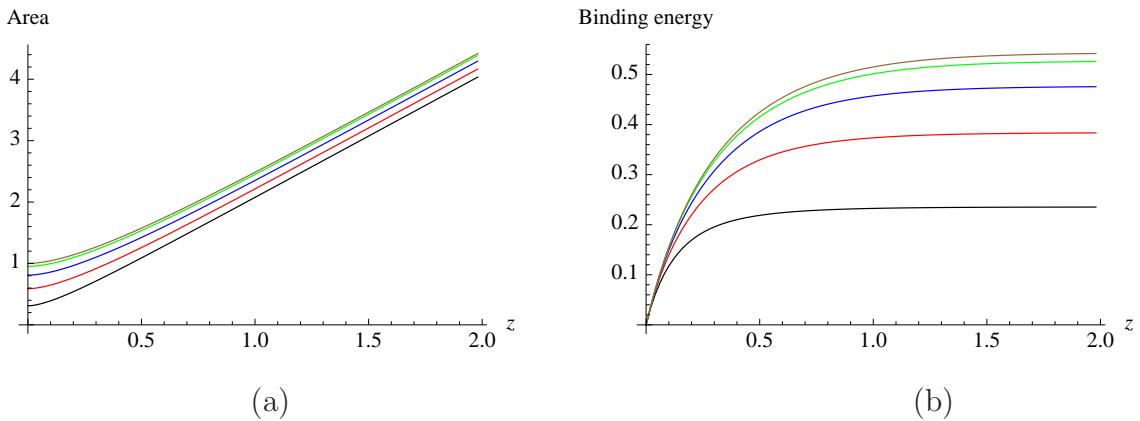}\\
 \hspace{1cm} (a) \hspace{8cm} (b)
\end{center}
 \vspace{-.5cm}
\caption{\sl \small (a) Minimal surface area and (b) the binding energy of the 't Hooft-Polyakov 
and Nambu monopoles  
for $\theta = \frac{\pi}{20},\ \frac{\pi}{10},\ \frac{3\pi}{20},\ \frac{\pi}{5},\ \frac{\pi}{4}$.
(a) The area of the minimal surface increases linearly when the 't Hooft-Polyakov monopole is much bigger than the Nambu monopole $(z \gg 1)$. (b) The binding energies reach constant values in any cases. 
}\label{result}
\end{figure}

We numerically compute the minimal surface area of the D2-brane and the binding energy of the Nambu
and 't Hooft-Polyakov monopoles. The results are shown in Fig.~\ref{result}.
As one can expect, 
the area of the minimal surface increases linearly when the 't Hooft-Polyakov monopole is much bigger
than the Nambu monopole $(z \gg 1)$. Also, the binding energies reach constant values in any cases.

\section{Discussions and open problems} \label{sec:summary}


In the previous section, we have argued that the electroweak string is stable when the adjoint field gets non-zero vacuum expectation value. In this case, the mass of the W-boson has to get a contribution from the VEV as in Eq.~(\ref{eq:ZWmass}). It is, however, not allowed from the precise current experimental  data. So in the true vacuum, the VEV of the adjoint field has to vanish. One interesting possibility is to consider the early Universe where the temperature is non-zero. In this case, thermal effect contributes to the potential and would change the stable point away from the origin. Suppose that thermal bath breaks the R-symmetry. In this case, the effective R-breaking effect would shift the minimum of the potential. Naively, we expect that the finite temperature effect to the moduli field should be given by 
\begin{equation}
V=(\Phi-a)^2 T^2+\cdots. 
\end{equation}

There is a possibility that a short-lived metastable string is created at high temperature and can be detected at 
colliders such as the Large Hadron Collider. It would be interesting to explore further on this avenue.


It would be interesting to consider a possible dual description of our model. According to \cite{GiveonKutasov}, the replacement of the NS5$_1$ and NS5$_2$ branes yields a dual description of the original model which can be interpreted as a Seiberg dual in field theory. In our setup, since $N_c=1$ and $N_f=2$, the dual description is self-dual and has the same gauge symmetry $U(N=N_f-N_c=1)$. In the dual description there exists a similar kind of string in a vacuum. It would be interesting to explore further in this direction and consider confining string by extending our study to the non-Abelian color and flavor symmetries. 
Such a Seiberg duality with non-Abelian vortices (without a monopole) 
in non-Abelian gauge theory 
was studied by replacing NS5 branes in the ${\cal N}=2$ context  \cite{Eto:2007yv}.
Our understanding of the monopole-string complex should play an important role. See \cite{Kitano} for a recent approach in this direction.


Realization of a non-BPS composite by a single bent D-brane can have straightforward extension to various dimensional objects. A domain wall ending on a string or bounded by a string loop is one of such examples. 
To this end, domain walls (without strings) in non-Abelian gauge theories 
\cite{Isozumi:2004jc} were realized by
D-brane configurations in type IIA/IIB string theory \cite{Eto:2004vy}.


\section*{Acknowledgments}

We would like to thank S.~Sugimoto, S.~Terashima and S.~Kanemura for discussions. M.~E., M.~N. and Y.~O. would like to thank the University of Pisa for their hospitality where this work was initiated. 
The work of M.~E. is supported by a 
Grant-in-Aid for Scientific Research from the Ministry of Education, Culture, Sports, Science and Technology, Japan (No. 23740226)
and the Japan Society for the Promotion of Science (JSPS) and Academy of Sciences of the Czech Republic (ASCR) 
under the Japan - Czech Republic Research Cooperative Program.
The work of M.~N. is supported in part by a 
Grant-in Aid for Scientific Research (No. 23740198) 
and by the ``Topological Quantum Phenomena'' 
Grant-in Aid for Scientific Research 
on Innovative Areas (No. 23103515)  
from the Ministry of Education, Culture, Sports, Science and Technology 
(MEXT) of Japan. 
Y.~O.'s research is supported by The Hakubi Center for Advanced Research, Kyoto University.

\section*{Appendix A\, Stabilization of moduli away from the origin }

In the main text, we mentioned ways to stabilize the moduli away from the origin. In the original model, the one-loop Coleman-Weinberg potential lifts the moduli at the origin. The generated mass term is of order ${\cal O}(h^2 \mu)$. 

Here we will show two possibilities to deform the theory for the stable point to be away from the origin. One is to add higher order of K\"ahler potential. The following correction term exists, 
\begin{equation}
K=|x-x_0|^2-{|x-x_0|^4 \over \Lambda^2}.
\end{equation}
The potentials of the theory is modified as follows:
\begin{equation}
V=|h\mu^2 |^2 K_{xx^*}^{-1}={|h\mu|^2  \over 1-{|x-x_0|^2 \over \Lambda^2}} \sim  {|h\mu^2 |^2 \over \Lambda^2}    |x-x_0|^2.
\end{equation}
Thus, as long as the condition
\begin{equation}
h < {\mu \over \Lambda}
\end{equation}
is satisfied, the correction terms coming from the K\"ahler potential are bigger than the one-loop Coleman-Weinberg potential. Stable point is around $x_0$. 

The other way to deform the theory is to break the R-symmetry by adding a quadratic term to the superpotential \cite{KOO, IntriligatorSeiberg}. For example, we add $W=\epsilon m x^2$ where $\epsilon$ is small parameter. In this case, by the balance with the Coleman-Weinberg potential, the minimum is shifted by the order ${\cal O}(\epsilon)$,
\begin{equation}
V= (h^2 \mu)^2 x^2 + \epsilon m h^2\mu^2 x +\cdots = (h^2 \mu^2)^2 (x + { \epsilon m \over h^2} )^2+\cdots.
\end{equation}

%
%


\begin{thebibliography}{1}
\bibitem{Higgs-mass}
  S.~Chatrchyan {\it et al.}  [CMS Collaboration],
  ``Observation of a new boson at a mass of 125 GeV with the CMS experiment at the LHC,''  Phys.\ Lett.\ B {\bf 716}, 30 (2012)  [arXiv:1207.7235 [hep-ex]];  
  G.~Aad {\it et al.}  [ATLAS Collaboration],
  ``Observation of a new particle in the search for the Standard Model Higgs boson with the ATLAS detector at the LHC,''  Phys.\ Lett.\ B {\bf 716}, 1 (2012)  [arXiv:1207.7214 [hep-ex]].  


\bibitem{Kibble:1976sj} 
  T.~W.~B.~Kibble,
  ``Topology of Cosmic Domains and Strings,''  J.\ Phys.\ A {\bf 9}, 1387 (1976).  

\bibitem{Zurek:1985}
W. H. Zurek,
``Cosmological experiments in superfluid helium?,"
Nature {\bf 317}, 505-508 (1985);
``Cosmological experiments in condensed matter systems,"
Phys. Rep. {\bf 276}, 177-221 (1996).

\bibitem{KZ-cond}
P. C. Hendry,
N. S. Lawson, R. A. M. Lee, P. V. E. Mcclintock, and C. D. H. Williams,
``Generation of defects in superfluid 4He as an analogue of the formation of cosmic strings,"
Nature {\bf 368}, 315-317 (1994);
M.~J.~Bowick, L. Chandar, E. A. Schiff and Ajit M. Srivastava,
``The Cosmological Kibble Mechanism in the Laboratory: String Formation in Liquid Crystals,"
Science {\bf 263}, 943-945 (1994);
C. B\"{a}uerle, Yu. M. Bunkov, S. N. Fisher, H. Godfrin, G. R. Pickett,
``Laboratory simulation of cosmic string formation in the early Universe using superfluid $^3$He," 
Nature {\bf 382}, 332-334 (1996);
V. M. H. Ruutu, V. B. Eltsov, A. J. Gill, T. W. B. Kibble, M. Krusius, Yu. G. Makhlin, B. Placais, G. E. Volovik,and Wen Xu,
``Vortex formation in neutron-irradiated superfluid 3He as an analogue of cosmological defect formation,"
Nature {\bf 382}, 334-336 (1996);
R. Carmi, E. Polturak, and G. Koren,
``Observation of Spontaneous Flux Generation in a Multi-Josephson-Junction Loop,"
Phys. Rev. Lett. {\bf 84}, 4966-4969 (2000);
A. Maniv, E. Polturak, and G. Koren,
``Observation of Magnetic Flux Generated Spontaneously During a Rapid Quench of Superconducting Films,"
Phys. Rev. Lett. {\bf 91}, 197001 (2003);
R. Monaco, J. Mygind, M. Aaroe, R. J. Rivers, and V. P. Koshelets,
``Zurek-Kibble Mechanism for the Spontaneous Vortex Formation in Nb-Al/Al$_{\rm ox}$/Nb Josephson Tunnel Junctions: New Theory and Experiment,"
Phys. Rev. Lett. {\bf 96}, 180604 (2006);
L. E. Sadler, J. M. Higbie, S. R. Leslie, M. Vengalattore, and D. M. Stamper-Kurn,
``Spontaneous symmetry breaking in a quenched ferromagnetic spinor Bose-Einstein condensate,"
 Nature {\bf 443}, 312-315 (2006);
Chad N. Weiler, Tyler W. Neely, David R. Scherer, Ashton S. Bradley, Matthew J. Davis, and Brian P. Anderson,
``Spontaneous vortices in the formation of Bose-Einstein condensates,"
 Nature {\bf 455}, 948-951 (2008).


\bibitem{Manton:1983nd} 
  N.~S.~Manton,
  ``Topology in the Weinberg-Salam Theory,''  Phys.\ Rev.\ D {\bf 28}, 2019 (1983);  
  F.~R.~Klinkhamer and N.~S.~Manton,
  ``A Saddle Point Solution in the Weinberg-Salam Theory,''  Phys.\ Rev.\ D {\bf 30}, 2212 (1984).  

\bibitem{Nambu} 
  Y.~Nambu,
  ``String-Like Configurations in the Weinberg-Salam Theory,''
  Nucl.\ Phys.\ B {\bf 130}, 505 (1977).

\bibitem{Vachaspati:1994xe} 
  T.~Vachaspati,
  ``Electroweak dyons,''  Nucl.\ Phys.\ B {\bf 439}, 79 (1995)  [hep-ph/9405285].  

\bibitem{Cho:1996qd} 
  Y.~M.~Cho and D.~Maison,
  ``Monopoles in Weinberg-Salam model,''  Phys.\ Lett.\ B {\bf 391}, 360 (1997)  [hep-th/9601028].  

\bibitem{Vachaspati:1992fi} 
  T.~Vachaspati,
  ``Vortex solutions in the Weinberg-Salam model,''  Phys.\ Rev.\ Lett.\  {\bf 68}, 1977 (1992)  [Erratum-ibid.\  {\bf 69}, 216 (1992)];  
  T.~Vachaspati,
  ``Electroweak strings,''  Nucl.\ Phys.\ B {\bf 397}, 648 (1993).  

\bibitem{JamesII} 
  M.~James, L.~Perivolaropoulos and T.~Vachaspati,
  ``Stability of electroweak strings,''
  Phys.\ Rev.\ D {\bf 46}, 5232 (1992).

\bibitem{JamesI} 
  M.~James, L.~Perivolaropoulos and T.~Vachaspati,
  ``Detailed stability analysis of electroweak strings,''
  Nucl.\ Phys.\ B {\bf 395}, 534 (1993)
  [hep-ph/9212301].



\bibitem{Goodband:1995he} 
  M.~Goodband and M.~Hindmarsh,
  ``Instabilities of electroweak strings,''  Phys.\ Lett.\ B {\bf 363}, 58 (1995)  [hep-ph/9505357].  


\bibitem{AVreview}
  A.~Achucarro and T.~Vachaspati,
  ``Semilocal and electroweak strings,''
  Phys.\ Rept.\  {\bf 327}, 347 (2000)
  [Phys.\ Rept.\  {\bf 327}, 427 (2000)]
  [arXiv:hep-ph/9904229].


\bibitem{Vachaspati:1994xc} 
  T.~Vachaspati,
  ``Electroweak strings, sphalerons and magnetic fields,''  hep-ph/9405286.  

\bibitem{Yang:1997qz} 
  Y.~Yang,
  ``Topological solitons in the Weinberg-Salam theory,''  Physica D {\bf 101}, 55 (1997).  

\bibitem{VA}
  T.~Vachaspati and A.~Achucarro,
  ``Semilocal cosmic strings,''
  Phys.\ Rev.\  D {\bf 44}, 3067 (1991).

\bibitem{Gibbons}
  G.~W.~Gibbons, M.~E.~Ortiz, F.~Ruiz Ruiz and T.~M.~Samols,
  ``Semilocal Strings And Monopoles,''
  Nucl.\ Phys.\  B {\bf 385}, 127 (1992)
  [arXiv:hep-th/9203023].

\bibitem{Hind}
  M.~Hindmarsh,
  ``Semilocal topological defects,''
  Nucl.\ Phys.\  B {\bf 392}, 461 (1993)
  [arXiv:hep-ph/9206229].

\bibitem{Vachaspati:1992mk} 
  T.~Vachaspati and R.~Watkins,
  ``Bound states can stabilize electroweak strings,''  Phys.\ Lett.\ B {\bf 318}, 163 (1993)  [hep-ph/9211284];  
  M.~A.~Earnshaw and W.~B.~Perkins,
  ``Stability of an electroweak string with a fermion condensate,''  Phys.\ Lett.\ B {\bf 328}, 337 (1994)  [hep-ph/9402218];  
  J.~M.~Moreno, D.~H.~Oaknin and M.~Quiros,
  ``Fermions on the electroweak string,''  Phys.\ Lett.\ B {\bf 347}, 332 (1995)  [hep-ph/9411411];  
  S.~G.~Naculich,
  ``Fermions destabilize electroweak strings,''  Phys.\ Rev.\ Lett.\  {\bf 75}, 998 (1995)  [hep-ph/9501388];  
  H.~Liu and T.~Vachaspati,
  ``Perturbed electroweak strings and fermion zero modes,''  Nucl.\ Phys.\ B {\bf 470}, 176 (1996)  [hep-ph/9511216];  
  M.~Groves and W.~B.~Perkins,
  ``The Dirac sea contribution to the energy of an electroweak string,''  Nucl.\ Phys.\ B {\bf 573}, 449 (2000)  [hep-ph/9908416];  
  G.~D.~Starkman, D.~Stojkovic and T.~Vachaspati,
  ``Neutrino zero modes on electroweak strings,''  Phys.\ Rev.\ D {\bf 63}, 085011 (2001)  [hep-ph/0007071];  
  G.~Starkman, D.~Stojkovic and T.~Vachaspati,
  ``Zero modes of fermions with a general mass matrix,''  Phys.\ Rev.\ D {\bf 65}, 065003 (2002)  [hep-th/0103039];  
  D.~Stojkovic,
  ``Neutrino zero modes and stability of electroweak strings,''  Int.\ J.\ Mod.\ Phys.\ A {\bf 16S1C}, 1034 (2001)  [hep-th/0103216];  
  N.~Graham, M.~Quandt and H.~Weigel,
  ``Fermion Energies in the Background of a Cosmic String,''  Phys.\ Rev.\ D {\bf 84}, 025017 (2011)  [arXiv:1105.1112 [hep-th]].  


\bibitem{Brandenberger:1994bx} 
  R.~H.~Brandenberger, A.~-C.~Davis and M.~Trodden,
  ``Cosmic strings and electroweak baryogenesis,''  Phys.\ Lett.\ B {\bf 335}, 123 (1994)  [hep-ph/9403215]; 
  A.~-C.~Davis, R.~H.~Brandenberger and M.~Trodden,
  ``Electroweak baryogenesis with topological defects,''  hep-ph/9406355.  

\bibitem{Poltis:2010yu} 
  R.~Poltis and D.~Stojkovic,
  ``Can primordial magnetic fields seeded by electroweak strings cause an alignment of quasar axes on cosmological scales?,''  Phys.\ Rev.\ Lett.\  {\bf 105}, 161301 (2010)  [arXiv:1004.2704 [astro-ph.CO]].  

\bibitem{Nambu:1974zg} 
  Y.~Nambu,
  ``Strings, Monopoles and Gauge Fields,''  Phys.\ Rev.\ D {\bf 10}, 4262 (1974);  
  S.~Mandelstam,
  ``Vortices And Quark Confinement In Nonabelian Gauge Theories,''  Phys.\ Lett.\ B {\bf 53}, 476 (1975);  
  S.~Mandelstam,
  ``Vortices and Quark Confinement in Nonabelian Gauge Theories,''  Phys.\ Rept.\  {\bf 23}, 245 (1976).  


\bibitem{Auzzi:2003fs} 
  R.~Auzzi, S.~Bolognesi, J.~Evslin, K.~Konishi and A.~Yung,
  ``NonAbelian superconductors: Vortices and confinement in N=2 SQCD,''  Nucl.\ Phys.\ B {\bf 673}, 187 (2003)  [hep-th/0307287].  

\bibitem{Auzzi:2003em} 
  R.~Auzzi, S.~Bolognesi, J.~Evslin and K.~Konishi,
  ``NonAbelian monopoles and the vortices that confine them,''  Nucl.\ Phys.\ B {\bf 686}, 119 (2004)  [hep-th/0312233];  
  M.~Eto, L.~Ferretti, K.~Konishi, G.~Marmorini, M.~Nitta, K.~Ohashi, W.~Vinci and N.~Yokoi,
  ``Non-Abelian duality from vortex moduli: A Dual model of color-confinement,''  Nucl.\ Phys.\ B {\bf 780}, 161 (2007)  [hep-th/0611313].  


\bibitem{KeVa} 
  T.~W.~Kephart and T.~Vachaspati,
  ``Topological incarnations of electroweak defects,''
  Phys.\ Lett.\ B {\bf 388}, 481 (1996)
  [hep-ph/9503355].

\bibitem{Aoki:2012yt} 
  M.~Aoki, S.~Kanemura, M.~Kikuchi and K.~Yagyu,
  ``Renormalization of the Higgs Sector in the Triplet Model,''
  Phys.\ Lett.\ B {\bf 714}, 279 (2012)
  [arXiv:1204.1951 [hep-ph]].

\bibitem{Hanany:2004ea} 
  A.~Hanany and D.~Tong,
  ``Vortex strings and four-dimensional gauge dynamics,''  JHEP {\bf 0404}, 066 (2004)  [hep-th/0403158].  

\bibitem{Auzzi:2004yg} 
  R.~Auzzi, S.~Bolognesi and J.~Evslin,
  ``Monopoles can be confined by 0, 1 or 2 vortices,''  JHEP {\bf 0502}, 046 (2005)  [hep-th/0411074].  


\bibitem{Abrikosov:1956sx} 
  A.~A.~Abrikosov,
  ``On the Magnetic properties of superconductors of the second group,''  Sov.\ Phys.\ JETP {\bf 5}, 1174 (1957)  [Zh.\ Eksp.\ Teor.\ Fiz.\  {\bf 32}, 1442 (1957)]; 
  H.~B.~Nielsen and P.~Olesen,
  ``Vortex Line Models for Dual Strings,''  Nucl.\ Phys.\ B {\bf 61}, 45 (1973).  

\bibitem{GiveonKutasov} 
  A.~Giveon and D.~Kutasov,
  ``Brane dynamics and gauge theory,''
  Rev.\ Mod.\ Phys.\  {\bf 71}, 983 (1999)
  [hep-th/9802067].

\bibitem{Kutasov1} 
  A.~Giveon and D.~Kutasov,
  ``Gauge Symmetry and Supersymmetry Breaking From Intersecting Branes,''
  Nucl.\ Phys.\ B {\bf 778}, 129 (2007)
  [hep-th/0703135 [hep-th]].

\bibitem{ISS} 
  K.~A.~Intriligator, N.~Seiberg and D.~Shih,
  ``Dynamical SUSY breaking in meta-stable vacua,''
  JHEP {\bf 0604}, 021 (2006)
  [hep-th/0602239].

\bibitem{Preskill} 
  J.~Preskill and A.~Vilenkin,
  ``Decay of metastable topological defects,''
  Phys.\ Rev.\ D {\bf 47}, 2324 (1993)
  [hep-ph/9209210].

\bibitem{BraneI}
  H.~Ooguri and Y.~Ookouchi,
  ``Meta-stable supersymmetry breaking vacua on intersecting branes,''
  Phys.\ Lett.\ B {\bf 641}, 323 (2006)
  [arXiv:hep-th/0607183].

\bibitem{BraneII}
  S.~Franco, I.~Garcia-Etxebarria and A.~M.~Uranga,
  ``Non-supersymmetric meta-stable vacua from brane configurations,''  JHEP {\bf 0701}, 085 (2007)  [hep-th/0607218].  

\bibitem{Bena:2006rg} 
  I.~Bena, E.~Gorbatov, S.~Hellerman, N.~Seiberg and D.~Shih,
  ``A Note on (Meta)stable Brane Configurations in MQCD,''  JHEP {\bf 0611}, 088 (2006)  [hep-th/0608157].  

\bibitem{KOOreview} 
  R.~Kitano, H.~Ooguri and Y.~Ookouchi,
  ``Supersymmetry Breaking and Gauge Mediation,''
  Ann.\ Rev.\ Nucl.\ Part.\ Sci.\  {\bf 60}, 491 (2010)
  [arXiv:1001.4535 [hep-th]].

\bibitem{EHT}
  M.~Eto, K.~Hashimoto and S.~Terashima,
  ``Solitons in supersymmety breaking meta-stable vacua,''
  JHEP {\bf 0703}, 061 (2007)
  [arXiv:hep-th/0610042].

\bibitem{IbeOokouchi} 
  K.~Hanaki, M.~Ibe, Y.~Ookouchi and C.~S.~Park,
  ``Constraints on Direct Gauge Mediation Models with Complex Representations,''
  JHEP {\bf 1108}, 044 (2011)
  [arXiv:1106.0551 [hep-ph]].

\bibitem{Tong}
  A.~Hanany and D.~Tong,
  ``Vortices, instantons and branes,''
  JHEP {\bf 0307}, 037 (2003)
  [arXiv:hep-th/0306150].

\bibitem{TongII}
  D.~Tong,
  ``TASI lectures on solitons,''
  arXiv:hep-th/0509216.

\bibitem{Eto:2005yh} 
  M.~Eto, Y.~Isozumi, M.~Nitta, K.~Ohashi and N.~Sakai,
  ``Moduli space of non-Abelian vortices,''  Phys.\ Rev.\ Lett.\  {\bf 96}, 161601 (2006)  [hep-th/0511088];  
  M.~Eto, Y.~Isozumi, M.~Nitta, K.~Ohashi and N.~Sakai,
  ``Solitons in the Higgs phase: The Moduli matrix approach,''  J.\ Phys.\ A A {\bf 39}, R315 (2006)  [hep-th/0602170];  
  M.~Eto, K.~Konishi, G.~Marmorini, M.~Nitta, K.~Ohashi, W.~Vinci and N.~Yokoi,
  ``Non-Abelian Vortices of Higher Winding Numbers,''  Phys.\ Rev.\ D {\bf 74}, 065021 (2006)  [hep-th/0607070];  
  M.~Eto, K.~Hashimoto, G.~Marmorini, M.~Nitta, K.~Ohashi and W.~Vinci,
  ``Universal Reconnection of Non-Abelian Cosmic Strings,''  Phys.\ Rev.\ Lett.\  {\bf 98}, 091602 (2007)  [hep-th/0609214].  

\bibitem{Eto:2007yv} 
  M.~Eto, J.~Evslin, K.~Konishi, G.~Marmorini, M.~Nitta, K.~Ohashi, W.~Vinci and N.~Yokoi,
  ``On the moduli space of semilocal strings and lumps,''  Phys.\ Rev.\ D {\bf 76}, 105002 (2007)  [arXiv:0704.2218 [hep-th]].  

 \bibitem{Kitano} 
  R.~Kitano, M.~Nakamura and N.~Yokoi,
  ``Making confining strings out of mesons,''  Phys.\ Rev.\ D {\bf 86}, 014510 (2012)  [arXiv:1202.3260 [hep-ph]].  

\bibitem{KOO} 
  R.~Kitano, H.~Ooguri and Y.~Ookouchi,
  ``Direct Mediation of Meta-Stable Supersymmetry Breaking,''
  Phys.\ Rev.\ D {\bf 75}, 045022 (2007)
  [hep-ph/0612139].
 
\bibitem{IntriligatorSeiberg} 
  K.~A.~Intriligator and N.~Seiberg,
  ``Lectures on Supersymmetry Breaking,''
  Class.\ Quant.\ Grav.\  {\bf 24}, S741 (2007)
  [hep-ph/0702069].

\bibitem{Isozumi:2004jc} 
  Y.~Isozumi, M.~Nitta, K.~Ohashi and N.~Sakai,
  ``Construction of non-Abelian walls and their complete moduli space,''  Phys.\ Rev.\ Lett.\  {\bf 93}, 161601 (2004)  [hep-th/0404198]; 
  Y.~Isozumi, M.~Nitta, K.~Ohashi and N.~Sakai,
  ``Non-Abelian walls in supersymmetric gauge theories,''  Phys.\ Rev.\ D {\bf 70}, 125014 (2004)  [hep-th/0405194];  
  M.~Eto, Y.~Isozumi, M.~Nitta, K.~Ohashi, K.~Ohta, N.~Sakai and Y.~Tachikawa,
  ``Global structure of moduli space for BPS walls,''  Phys.\ Rev.\ D {\bf 71}, 105009 (2005)  [hep-th/0503033].  

\bibitem{Eto:2004vy} 
  M.~Eto, Y.~Isozumi, M.~Nitta, K.~Ohashi, K.~Ohta and N.~Sakai,
  ``D-brane construction for non-Abelian walls,''  Phys.\ Rev.\ D {\bf 71}, 125006 (2005)  [hep-th/0412024].  

 
\end{thebibliography}
\end{document}